\documentclass[10pt,a4paper]{article}
\usepackage{jheppub_kim}
\usepackage{pdflscape}
\usepackage{amsmath}
\usepackage{amssymb}
\usepackage{dcolumn}
\usepackage{bm}
\usepackage{color}
\usepackage{epsfig}
\usepackage{amsfonts}
\usepackage{graphicx}
\usepackage{subfigure}
\usepackage{dcolumn}

\begin{document}
\title{Non-perturbative correction to the Ho\v{r}ava-Lifshitz black hole thermodynamics}
\author[a]{Behnam Pourhassan,}
\author[b]{\.{I}zzet Sakall{\i}}

\affiliation[a] {School of Physics, Damghan University, Damghan, 3671641167, Iran.}
\affiliation[a] {Physics Department, Istanbul Technical University, Istanbul 34469, Turkey.}
\affiliation[a] {Canadian Quantum Research Center 204-3002, 32 Ave Vernon, BC V1T 2L7 Canada.}

\affiliation[b] {Physics Department, Eastern Mediterranean
University, Famagusta 99628, North Cyprus via Mersin 10, Turkey.}

\emailAdd{b.pourhassan@du.ac.ir} \emailAdd{izzet.sakalli@emu.edu.tr}

\abstract{In this paper, we consider non-perturbative quantum correction which appears as exponential term in the black hole entropy. We study consequence thermodynamics of the Ho\v{r}ava-Lifshitz black hole at quantum scales. We consider two cases of Kehagius-Sfetsos and Lu-Mei-Pop solutions and investigate black hole stability. We find that non-perturbative quantum correction yields to an instability at infinitesimal horizon radius of Kehagius-Sfetsos solution. On the other hand, non-perturbative quantum correction yields to the stability of Lu-Mei-Pop solution. Hence, we find that holographic dual of Lu-Mei-Pop black hole (in absence of non-perturbative quantum correction) is the interacting gas of point like particles, while it is Van der Waals fluid in presence of non-perturbative quantum correction.}


\maketitle

\section{Introduction}
One of the finest ways to learn about black holes is to study their thermodynamics. By utilizing the surface information, the holographic principle allows us to obtain insight into black hole physics. The black hole entropy is proportional to the event horizon area and the black hole temperature is proportional to the black hole's surface gravity, hence the principles of thermodynamics are satisfied \cite{0001,0002,0003}. If we consider the following general line element to describe a black hole:
\begin{equation}\label{0005}
ds^2=f(r)dt^{2} - \frac{dr^{2}}{f(r)} - r^{2}dX^{2},
\end{equation}
then the black hole temperature is obtained using the derivative of metric function $f(r)$ at the horizon radius. Also, horizon area (hence entropy) is obtained from $dX^{2}$ geometry. Having the black hole temperature and entropy, one can study the black hole thermodynamics \cite{0004, R16}. There are several kinds of black holes, the simplest one is described by only one parameter which is the black hole mass, it is Schwarzschild black hole \cite{0005, 0006}. A black hole carrying the electric charge is known as the Reissner Nordstr\"{o}m black hole and was already been examined from a thermodynamics standpoint \cite{0007}. Adding rotation to the Schwarzschild black hole yields the well-known Kerr black hole. Thermodynamics, stability, and Hawking-Page phase transition of the Kerr black hole were studied in Ref. \cite{0008}. The Kerr-Newman black holes are described by three parameters; mass, electric charge, and rotation, which govern the black hole thermodynamics \cite{0009}. On the other hand, a black hole in anti-de Sitter (AdS) space \cite{AdS} is significant in terms of gauge/gravity \cite{00010, 00011}. The thermodynamics of some AdS black holes have been studied where the cosmological constant plays the role of black hole pressure \cite{00012, 00013, 00014}.
Other fascinating types of black holes including Schr\"{o}dinger black holes \cite{00015, 00016, 00017, 00018} and stringy black holes \cite{00019} provide information regarding supergravity.\\

Ho\v{r}ava-Lifshitz black holes, which are the subject of this study, are also among the most fascinating types of black holes \cite{00020, 00021}. With general coupling constants, Ho\v{r}ava-Lifshitz black holes admit spherically symmetric solutions \cite{00022}. Those black holes are first constructed by Ho\v{r}ava \cite{00023, 00024}. Ho\v{r}ava-Lifshitz gravity is indeed a theory of quantum gravity, which is nothing but a renormalizable non-relativistic theory of gravity \cite{00025, 00026, 00027}. At large radii, Ho\v{r}ava-Lifshitz black hole solutions can be reduced to Einstein's general relativity black hole, whereas at small scales they reveal violated Lorentz symmetry \cite{00028}.\\

Thermodynamic analysis of black holes shows that their emission decreases as their mass decreases. When the mass and size of a black hole are reduced to the quantum scale, quantum corrections become more relevant. What happens to a black hole at quantum sizes is an important question. Black hole evaporation is a well-known phenomenon that has also been investigated in the Ho\v{r}ava-Lifshitz gravity model \cite{000281}. There are two kinds of perturbative and non-perturbative corrections. The thermal fluctuations are calculated via perturbative quantum field theory and they usually appear as a logarithmic correction term (at leading order) in the black hole entropy \cite{00029, 00030, 00031, 00032, JHAP, R1, R2, R3, R4, R5, R6, R7, R8, R9, R10, R11, R12, R13, R14, R15}. On the other hand, higher order perturbative corrections appear as power law \cite{00033, 00034, higher001}. It is worth noting that non-perturbative quantum effects can correct the black hole entropy by exponential term \cite{00035}. In that case, exponential correction on the thermodynamics of Schwarzschild-AdS, Reissner-Nordstr\"{o}m, and charged AdS black holes have been investigated recently by Ref. \cite{2010.03946} and Myers-Perry black holes by Ref. \cite{expJHEP00}. The non-perturbative corrections can connect topology to the quantum gravity \cite{log5}, which may be used to test the quantum gravity \cite{test003}. It should be also noted that the exponential correction is not only used for the black holes but also used for brane systems \cite{expJHEP01, expJHEP02}\\

In this paper, we want to study the exponential correction on the thermodynamics of Ho\v{r}ava-Lifshitz black hole to find the effect of non-perturbative corrections. The exponential correction can modify the black hole entropy, hence they may be effective only on the $dX^{2}$ geometry. So, the black hole temperature and horizon radius (metric function) are affected by non-perturbative corrections. We expect it to have an impact on black hole stability, allowing us to study critical points and phase transitions in Ho\v{r}ava-Lifshitz black holes at quantum scales.\\

The paper is organized as follows. In section \ref{Ec}, we introduce non-perturbative correction and obtain the exponential term of the black hole entropy. In section \ref{thermo}, we review important aspects of Ho\v{r}ava-Lifshitz black hole and consider two special cases. Corrected thermodynamics of the first and second cases are studied in sections \ref{KS} and \ref{LMP}, respectively. The stability of these cases is investigated in section \ref{stab}. In section \ref{EoS}, we study the effect of the non-perturbative quantum corrections on the state equation of the second case. Section \ref{con} is devoted to summary and conclusions.

\section{Exponential correction of the entropy}\label{Ec}
It is well-known that the entropy of a black hole is proportional to the horizon area. Classical black holes are stable objects which never vanish. A black hole, on the other hand, can emit what is known as Hawking radiation. It is from a quantum mechanical standpoint that the black hole's size is reduced. There are two possibilities for the final stage of a black hole. The first one is the evaporation of a black hole when all of its mass radiate. In the second scenario, the black hole stops its radiation at a specific stage and becomes stable. Quantum effects are to blame for these events. Thus, we should investigate black hole thermodynamics under quantum effects in order to uncover such possibilities. In that case, we may find why the Ho\v{r}ava-Lifshitz black holes become an effective remnant after the end of the evaporation \cite{000281}. These effects are important when the black hole size is infinitesimal compared to the Planck scale. There are also perturbative and non-perturbative quantum effects. The leading order of the perturbative corrections to the black hole entropy is usually logarithmic \cite{log0,22} or power law for the black holes in higher dimensions \cite{P1}. A similar situation happens for the higher order corrections \cite{higher3}. On the other hand, the black hole corrected entropy can be obtained by using the non-perturbative quantum gravity theory.\\

To obtain exponential correction, we consider a black hole involving a total $N$ particles, which are constructed by $n_{i}$ number. One can assume that each number $n_{i}$ is shared by $s_{i}$ tiles. Therefore, the total number is given by
\begin{equation}\label{St1}
N=\Sigma{s_{i}n_{i}}.
\end{equation}
Besides, the total number of microstates $(\Omega)$ of a black hole is given by
\begin{equation}\label{Sm}
\Omega=\frac{N!}{\Pi{n_{i}s_{i}}!},
\end{equation}
which can be rewritten as following
\begin{equation}\label{Sm-log}
\ln{\Omega}=\ln{N!}-\ln{\Pi{n_{i}s_{i}}!}.
\end{equation}
The most likely configuration is determined by varying $\ln{\Omega}$ \cite{2010.03946}:
\begin{equation}\label{S00}
s_{i}=\left(\sum{n_{i}}\right)e^{-\lambda n_{i}},
\end{equation}
where the parameter $\lambda$ is called the variation parameter, which plays the role of Lagrange multipliers. It satisfies the following condition \cite{2007.15401}:
\begin{equation}\label{lambda}
\sum{e^{-\lambda n_{i}}}=1,
\end{equation}
which yields
\begin{equation}\label{lambda1}
\lambda\approx\ln{2}-2^{-N},
\end{equation}
where $\mathcal{O}(2^{-2N})$ is neglected. In that case, the entropy is given by
\begin{equation}\label{en}
S=\lambda N.
\end{equation}

\begin{figure}[h!]
\begin{center}$
\begin{array}{cccc}
\includegraphics[width=50 mm]{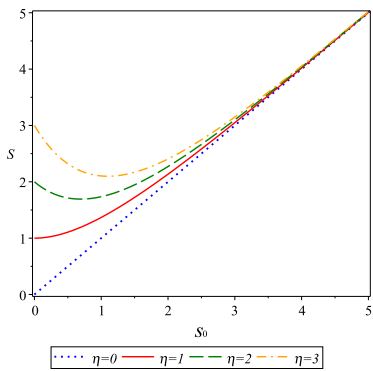}
\end{array}$
\end{center}
\caption{Corrected entropy in terms of the original entropy.}
\label{fig1}
\end{figure}

So, the combination of Eqs. (\ref{lambda1}) and (\ref{en}) suggests that the black hole entropy is corrected by an exponential term as following
\begin{equation}\label{S}
S=S_{0}+\eta e^{-S_{0}},
\end{equation}
where $S_{0}$ is the original entropy and $\eta$ denotes the correction parameter. The case of $\eta=1$ is discussed by Ref. \cite{2007.15401}, in which it was argued that this exponential corrected term acts as a non-perturbative quantum correction, which is independent of the quantum gravity theory. The correction parameter $\eta$ could be the difference between several quantum gravity theories. The same argument happens for the logarithmic correction of the black hole entropy \cite{EPL}. From Eq. (\ref{S}), we can see that $S=S_{0}$ for the large black hole where $A\gg1$, hence $S_{0}\gg1$. Therefore, this term is negligible for the large black holes (see Fig. \ref{fig1}), and it is considered as a non-perturbative quantum effect for the infinitesimal black hole similar to the perturbative corrections \cite{NPB}. An interesting point is that the total entropy is not zero when the horizon area vanishes. It might like two-dimensional black holes, where the horizon area is a point and the entropy can only be derived by using the entropy function approach \cite{Sen001,e1}.

\section{Ho\v{r}ava-Lifshitz black hole solutions}\label{thermo}
The four-dimensional Ho\v{r}ava-Lifshitz black hole is described by the following metric \cite{12, EPJC},
\begin{equation}\label{5}
ds^2=f(r)dt^2 - \frac{dr^2}{f(r)} - r^{2}d\Omega_{2}^{2},
\end{equation}
where
\begin{equation}\label{7}
f(r)=k + (\omega - \Lambda_W)r^{2} -\sqrt{r(\omega(\omega - 2\Lambda_W)r^{3} + B)},
\end{equation}
in which $B$ is an integration constant while $\omega$, $\Lambda_W$, and $k$ are constant parameters \cite{13}. The constant $\Lambda_W\leq0$ is related to the cosmological constant via $\Lambda=\frac{3}{2}\Lambda_W$ \cite{AS}. Also, $d\Omega_{2}^{2}$ is a two-dimensional spherical, flat, or hyperbolic space which is given by $k=1$, $k=0$, and $k=-1$, respectively.\\

\begin{figure}[h!]
\begin{center}$
\begin{array}{cccc}
\includegraphics[width=47 mm]{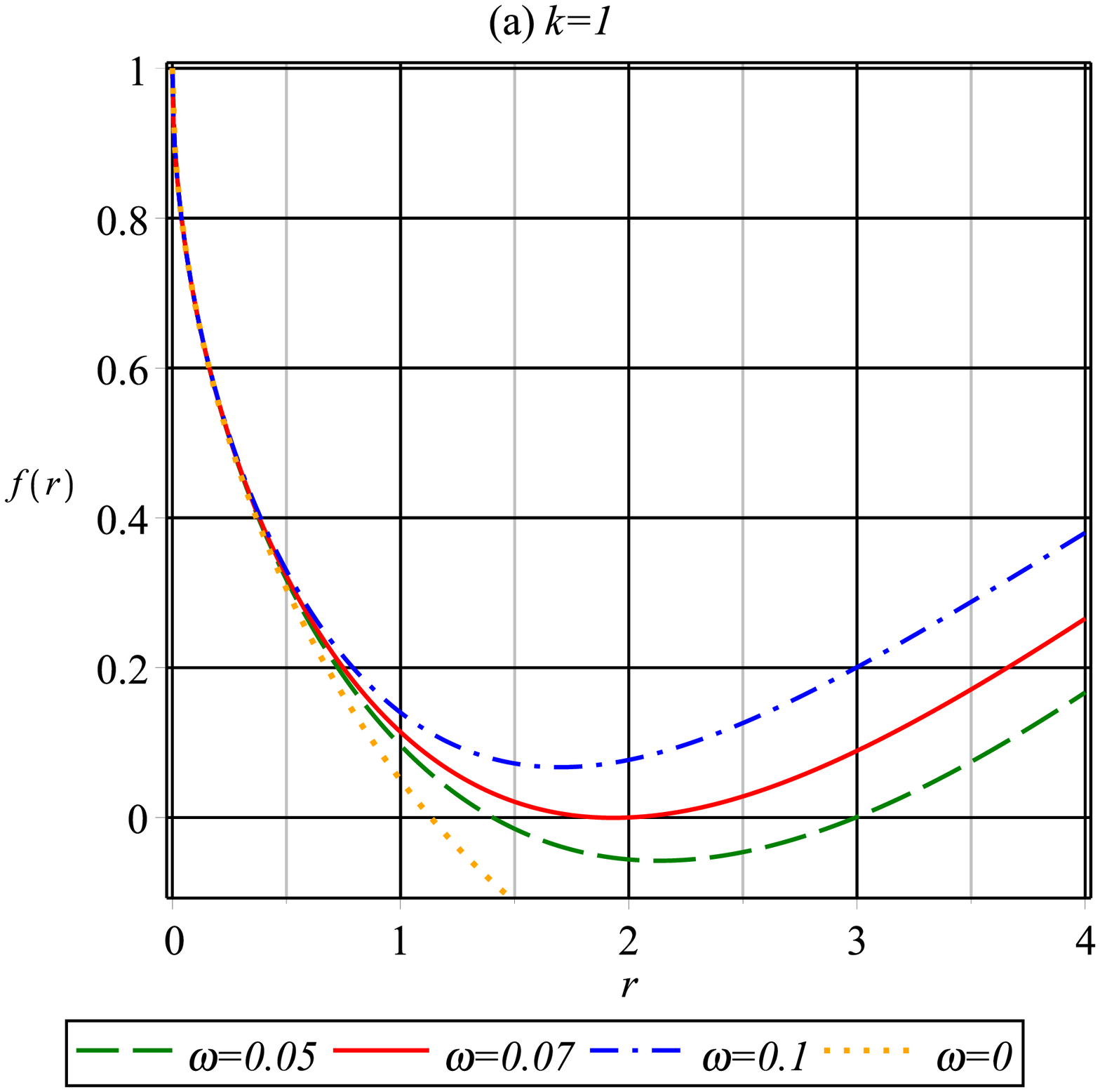}\includegraphics[width=47 mm]{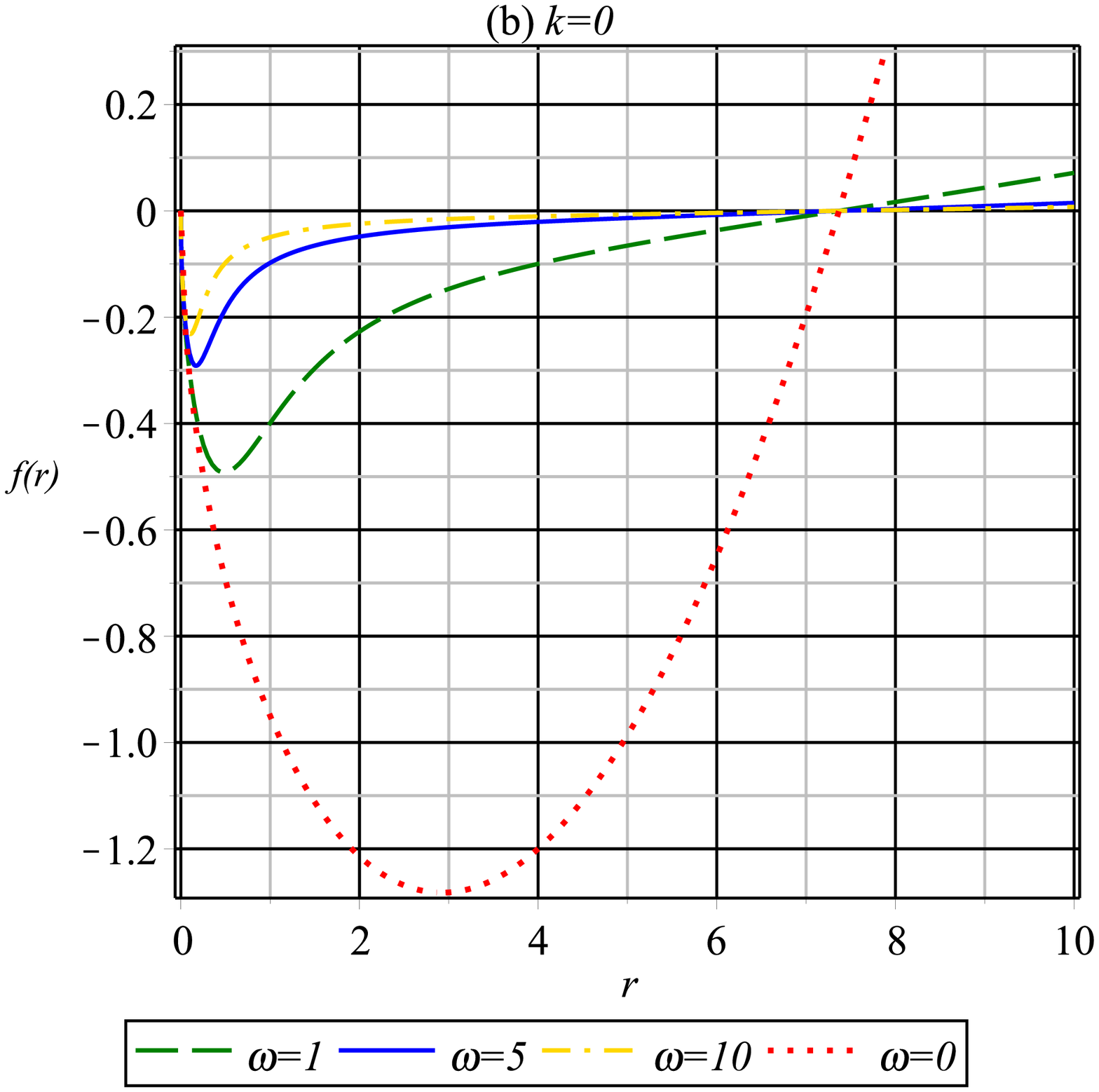}\includegraphics[width=47 mm]{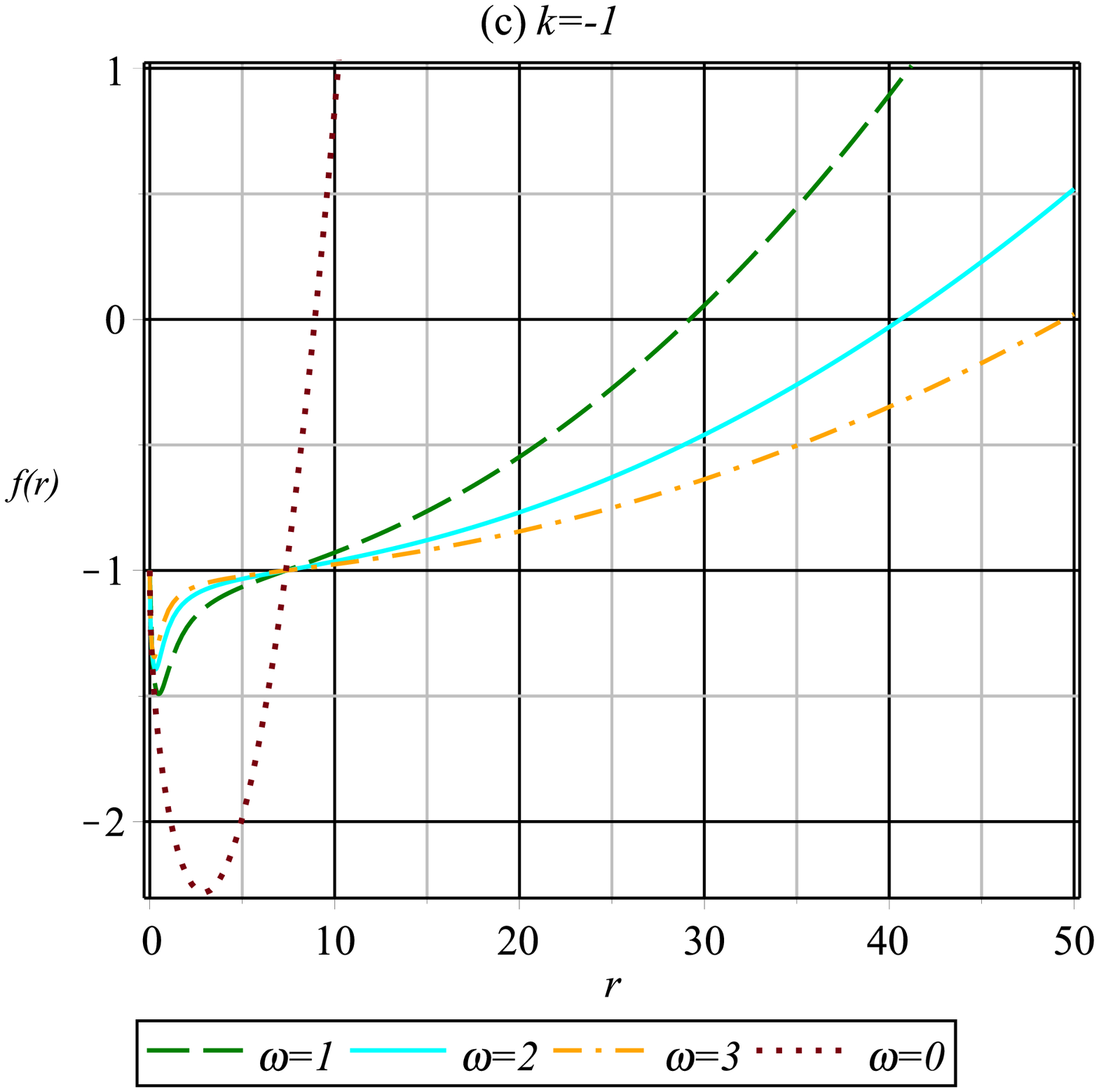}
\end{array}$
\end{center}
\caption{Horizon structures of Ho\v{r}ava-Lifshitz black holes with $\Lambda_W=-0.05$ and $B=1$.}
\label{fig2}
\end{figure}

In Fig. \ref{fig2}, one can see the horizon structures of Ho\v{r}ava-Lifshitz black holes for three different geometries: spherical ($k=1$), flat ($k=0$), and open ($k=-1$). In the case of $k=1$, there are three situations for the two distinct horizons ($r_{-}$ and $r_{+}$; see the dashed green line of Fig. \ref{fig2} (a), for the extremal case ($r_{-}=r_{+}$; see the solid red line of Fig. \ref{fig2} (a), and for the naked singularity case see the dash-dotted blue line of Fig. \ref{fig2} (a). Also, in Fig. \ref{fig2} (b) and (c), we can see that value of $r_{+}$ does not depend on $\omega$.\\

The Lagrangian density of corresponding action \cite{14,15,16} is given by,
\begin{equation}\label{01}
L=\sqrt{g}N\mathcal{L},
\end{equation}
where $g$ is determinant of metric (\ref{5}) and
\begin{eqnarray}\label{1}
\mathcal{L}&=&\frac{2}{\kappa^2}(K_{ij} K^{ij}-\lambda K^{2})+\frac{\kappa^2 \mu^2(\Lambda_W
R-3\Lambda_W^2)}{8(1-3\lambda)}+\frac{\kappa^2\mu^2(1-4\lambda)}{32(1-3\lambda)}R^{2}\nonumber\\
&-&\frac{\kappa^2\mu^2}{8}R_{ij}R^{ij}
 +\frac{\kappa^2\mu}{2\omega^2}\epsilon^{ijk} R_{il}\nabla_j
R_k^{l} - \frac{\kappa^2}{2\omega^4}C_{ij}C^{ij},
\end{eqnarray}
by which $\kappa$, $\lambda$, and $\mu$ represent the constant parameters, while $C_{ij}$ is the Cotton tensor given by \cite{14, 17-1},
\begin{equation}\label{2}
C^{ij}=\epsilon^{ikl} \nabla_k (R_{l}^{j} - \frac{1}{4} R\delta_{l}^{j}).
\end{equation}
On the other hand, $K_{ij}$ is the extrinsic curvature given by (for the time-independent metric function)
\begin{equation}\label{3}
K_{ij}=-\frac{1}{2N}(\nabla_i N_j + \nabla_j N_i ),
\end{equation}
where $N_i$ and $N$ are shift and lapse functions, respectively \cite{EPJC}. After making straightforward calculations, for all three cases of $k=1,0,-1$, one gets the following Ricci scalar:
\begin{equation}\label{Ricci}
R=-\frac{r^{2}f^{\prime\prime}(r)+4rf^{\prime}(r)+2f(r)+2}{r^{2}}.
\end{equation}
For metric \eqref{5}, the black hole temperature is given by \cite{wald}
\begin{equation}\label{temp}
T=\frac{1}{4\pi}\left(\frac{df(r)}{dr}\right)_{r=r_{+}}.
\end{equation}
In Fig. \ref{fig2}, the general structure for the Ho\v{r}ava-Lifshitz black holes is represented. However, there are two particular cases, which are going to be discussed in the following subsections.

\subsection{KS solution}
\begin{figure}[h!]
\begin{center}$
\begin{array}{cccc}
\includegraphics[width=60 mm]{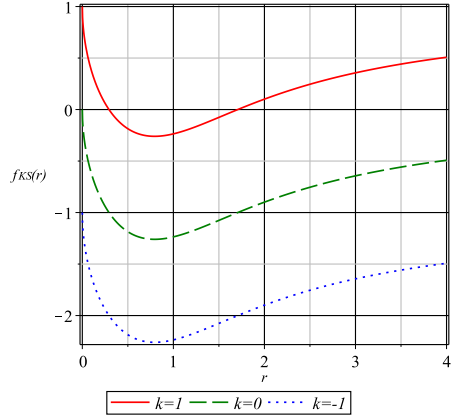}
\end{array}$
\end{center}
\caption{Horizon structures of KS solution of Ho\v{r}ava-Lifshitz black holes with $\omega=1$ and $m=1$.}
\label{fig3}
\end{figure}

\begin{figure}[h!]
\begin{center}$
\begin{array}{cccc}
\includegraphics[width=60 mm]{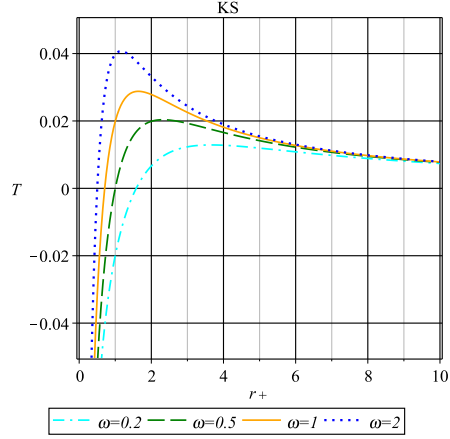}
\end{array}$
\end{center}
\caption{Temperature of KS solution of Ho\v{r}ava-Lifshitz black holes with $k=1$.}
\label{fig4}
\end{figure}

Kehagius-Sfetsos (KS) solution is one of the Ho\v{r}ava-Lifshitz black holes having zero cosmological constant with $B=4\omega M$ \cite{19}
\begin{equation}\label{8}
f_{KS}(r)=k +\omega r^{2}- \omega r^{2}\sqrt{1+\frac{4m}{\omega r^{3}}},
\end{equation}
where $m$ is the mass parameter so that black hole mass is given by $M=am$, where $a$ is a positive constant. Using $f_{KS}(r)=0$, we find inner ($r_{-}$) and outer ($r_{+}$) horizons as
\begin{equation}\label{h}
r_{\pm}=\frac{2\omega m\pm\sqrt{4\omega^{2}-2\omega k^{3}}}{2k\omega}.
\end{equation}
In Fig. \ref{fig3}, one can see that values of horizon radius for the cases of $k=0$ and $k=-1$ are larger than the case of $k=1$. Hence, in order to discuss the non-perturbative features of the small black holes, we will only consider the case of spherical geometry.
In that case, Ricci scalar (\ref{Ricci}) yields
\begin{equation}\label{Ricci-KS}
R=\frac{2\left[6\omega^{4}r^{8}+36m\omega^{3}r^{5}-(k+6\omega r^{2})\left(\omega^{2}r^{4}+4m \omega r\right)^{\frac{3}{2}}+30m^{2}\omega^{2}r^{2}\right]}{\left(\omega^{2}r^{4}+4m \omega r\right)^{\frac{3}{2}}r^{2}}.
\end{equation}
As being discussed by Ref. \cite{16}, this case can be reduced to the Schwarzschild black hole. In the case of $k=1$, the mass parameter is obtained as
\begin{equation}\label{mass-KS}
m=\frac{1+2\omega r_{+}^{2}}{4\omega r_{+}}.
\end{equation}
Using the mass parameter (\ref{mass-KS}) and expression (\ref{temp}), we can find the KS black hole temperature as following,
\begin{equation}\label{temp-KS}
T=\frac{2\omega r_{+}^{2}-1}{8\pi r_{+}(1+\omega r_{+}^{2})}.
\end{equation}
The above temperature equation demonstrates that the situation of $\omega=0$ results in negative temperature, which is clearly not a physical case. For the cases of $\omega\neq0$, one can find a minimum value for the event horizon radius, which is henceforth symbolized by $r_{+m}$. So for the cases of $r_{+}\leq r_{+m}$, we have $T\leq0$, which is again a nonphysical situation. From Eq. (\ref{temp-KS}), it is easy to get that
\begin{equation}\label{rm}
r_{+m}^{2}=\frac{1}{2\omega},
\end{equation}
which yields $T(r_{+}=r_{+m})=0$. Hence, the value of $r_{+m}$ depends on $\omega$. In Fig. \ref{fig4}, we can see the behavior of temperature and observe that there is a maximum for the critical value of the event horizon radius $r_{+c}$ which is given by
\begin{equation}\label{rc}
r_{+c}^{2}=\frac{5+\sqrt{33}}{4\omega}.
\end{equation}
A phase transition could be indicated by the maximum temperature. As a result, it merits more examination, which we will conduct later utilizing the specific heat analysis. Besides, at large $r_{+}$, the value of $\omega$ has no important effect on the temperature:
\begin{equation}\label{temp-KS-asymp}
T(r_{+}\rightarrow\infty)\approx\frac{1}{4\pi r_{+}}.
\end{equation}
Since the black hole pressure is proportional to the cosmological constant, there is no variable corresponding to the thermodynamics pressure of this case. Hence, the first law of black hole thermodynamics reads
\begin{equation}\label{first law KS}
dE=TdS,
\end{equation}
where $E$ is the internal energy and $S$ is the black hole entropy, which was given in Eq. (\ref{S}). To have more information about the black hole thermodynamics of KS solution, the interested reader is referred to Ref. \cite{AS}, which will be used by us to study the corrected thermodynamics in the next section.

\subsection{LMP solution}
\begin{figure}[h!]
\begin{center}$
\begin{array}{cccc}
\includegraphics[width=60 mm]{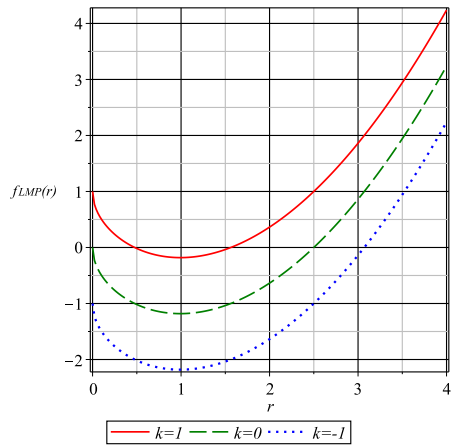}
\end{array}$
\end{center}
\caption{Horizon structures of LMP solution of Ho\v{r}ava-Lifshitz black holes with $\Lambda_W=-0.4$ and $m=1$.}
\label{fig5}
\end{figure}

After inserting $\omega=0$ and $B=-\frac{a^{2}m^{2}}{\Lambda_W}$ into the expression (\ref{7}), one gets Lu-Mei-Pop (LMP) solution \cite{12} which is given by
\begin{equation}\label{9}
f_{LMP}(r)=k - \Lambda_Wr^{2} - am \sqrt{\frac{r}{-\Lambda_W}},
\end{equation}
where $a$ is a constant parameter, which can be set to unity for simplicity. In Fig. \ref{fig5}, we can see that the smallest root belongs to the spherical geometry. Therefore similar to the previous subsection, we solely consider the case of $k=1$.\\

\begin{figure}[h!]
\begin{center}$
\begin{array}{cccc}
\includegraphics[width=55 mm]{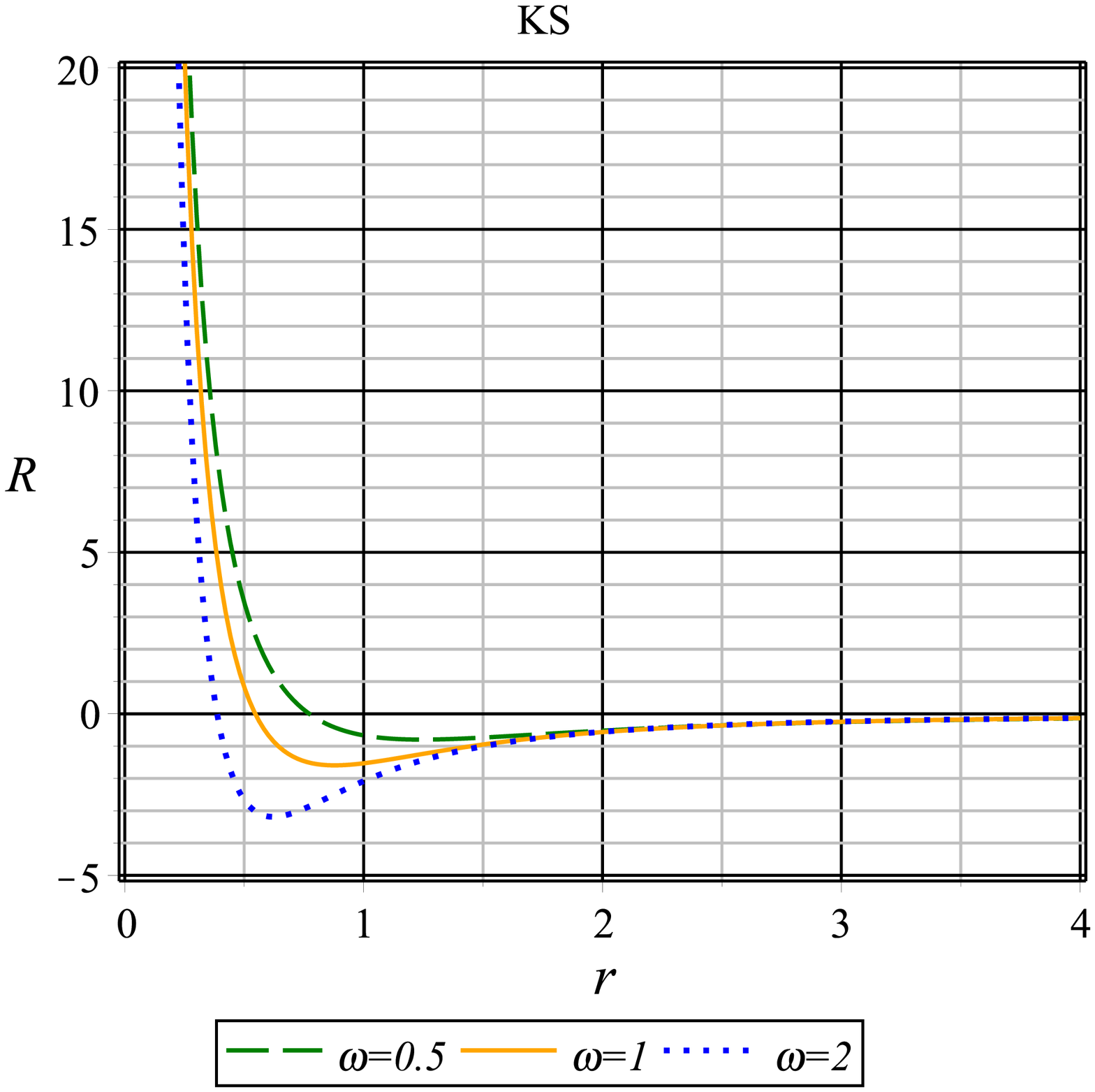}\includegraphics[width=55 mm]{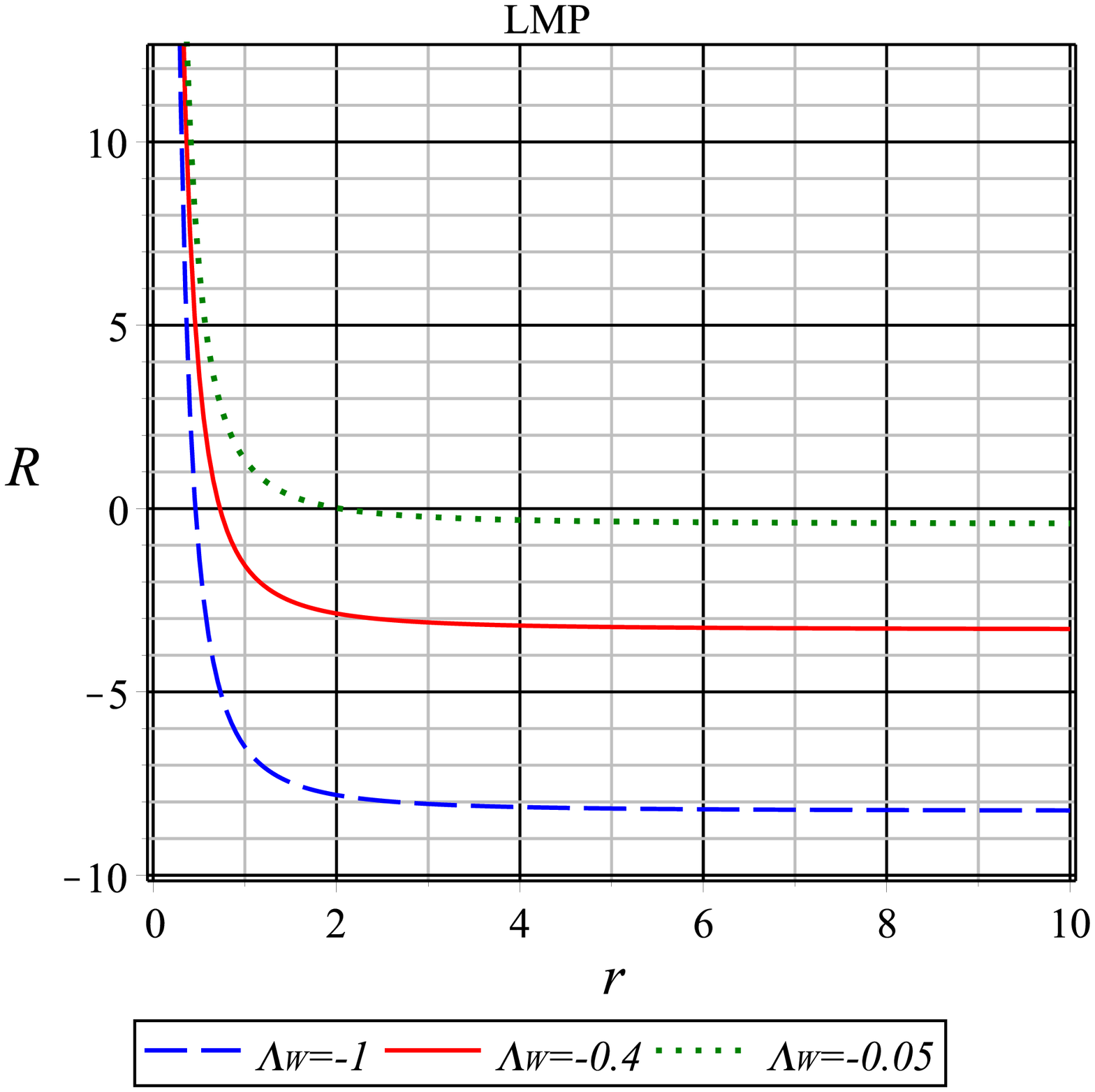}
\end{array}$
\end{center}
\caption{Ricci scalar of Ho\v{r}ava-Lifshitz black holes with $k=1$.}
\label{fig6}
\end{figure}

In the case of $k=1$, the mass parameter is given by
\begin{equation}\label{mass-LMP}
m=\frac{\sqrt{-\Lambda_W r_{+}}(\Lambda_W r_{+}^{2}-1)}{r_{+}}.
\end{equation}
Ricci scalar of this case is obtained as following
\begin{equation}\label{Ricci-LMP}
R=\frac{15m}{4\Lambda^{2}}\left(-\frac{\Lambda}{r}\right)^{\frac{3}{2}}-\frac{2}{r^{2}}(k-6\Lambda r^{2}).
\end{equation}
Eliminating the mass parameter from Eqs. (\ref{Ricci-KS}) and (\ref{Ricci-LMP}), one can exhibit the behavior of Ricci scalar of KS and LMP solutions: see Fig. \ref{fig6}.

Using the mass parameter (\ref{mass-LMP}) and Eq. (\ref{temp}), and further setting $k=1$, one can derive the LMP black hole temperature as
\begin{equation}\label{temp-LMP}
T=\frac{1-5\Lambda_{W}r_{+}^{2}}{8\pi r_{+}}.
\end{equation}
The above result shows that the temperature is completely positive for the negative cosmological constant, which is a physical situation. However, there is a minimum value for the temperature at the critical value of the event horizon radius, $r_{+c}$. From Eq. (\ref{temp-LMP}), it is easy to obtain,
\begin{equation}\label{rm-LMP}
r_{+c}^{2}=-\frac{1}{5\Lambda_{W}}.
\end{equation}
Moreover, it is clear that for small values of $r_{+}$, $\Lambda_{W}$ has no significant effect on the black hole temperature. As is known, the black hole pressure is proportional to the cosmological constant \cite{18-0}. Thus, we have
\begin{equation}\label{pressure}
P=-\frac{3\Lambda_{W}}{16\pi}.
\end{equation}
Therefore, the first law of black hole thermodynamics for this case reads
\begin{equation}\label{first law KS}
dE=TdS+VdP,
\end{equation}
where $E$ is the internal energy and $S$ is the black hole entropy (\ref{S}).
Thermodynamical quantities of the LMP solution was already studied by Refs. \cite{main, NPB2}. In the next sections, we shall discuss the corrected thermodynamics relations for the Ho\v{r}ava-Lifshitz black holes.

\section{Corrected thermodynamics of KS solution}\label{KS}
\begin{figure}[h!]
\begin{center}$
\begin{array}{cccc}
\includegraphics[width=55 mm]{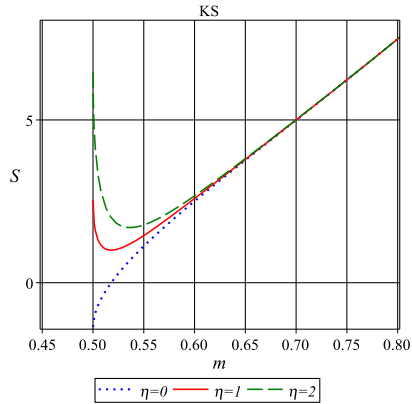}
\end{array}$
\end{center}
\caption{Corrected entropy of KS solution with $k=1$ and $\omega=2$.}
\label{fig7}
\end{figure}

Entropy of the KS black hole with $k=1$ is given by \cite{EPJC},
\begin{equation}\label{s-KS}
S_{0}=\pi r_{+}^{2}+\frac{2\pi}{\omega}\ln{r_{+}}.
\end{equation}
Hence, using Eq. (\ref{S}), one can write the exponentially corrected entropy as
\begin{equation}\label{S-KS}
S=\pi r_{+}^{2}+\frac{2\pi}{\omega}\ln{r_{+}}+\eta r_{+}^{-\frac{2\pi}{\omega}} e^{-\pi r_{+}^{2}},
\end{equation}
where $r_{+}$ is defined in Eq. (\ref{h}). In Fig. \ref{fig7}, for the special range of the black hole mass, we show the effect of the exponential correction on the black hole entropy, and thus reveal the importance of the non-perturbative effects.

\begin{figure}[h!]
\begin{center}$
\begin{array}{cccc}
\includegraphics[width=55 mm]{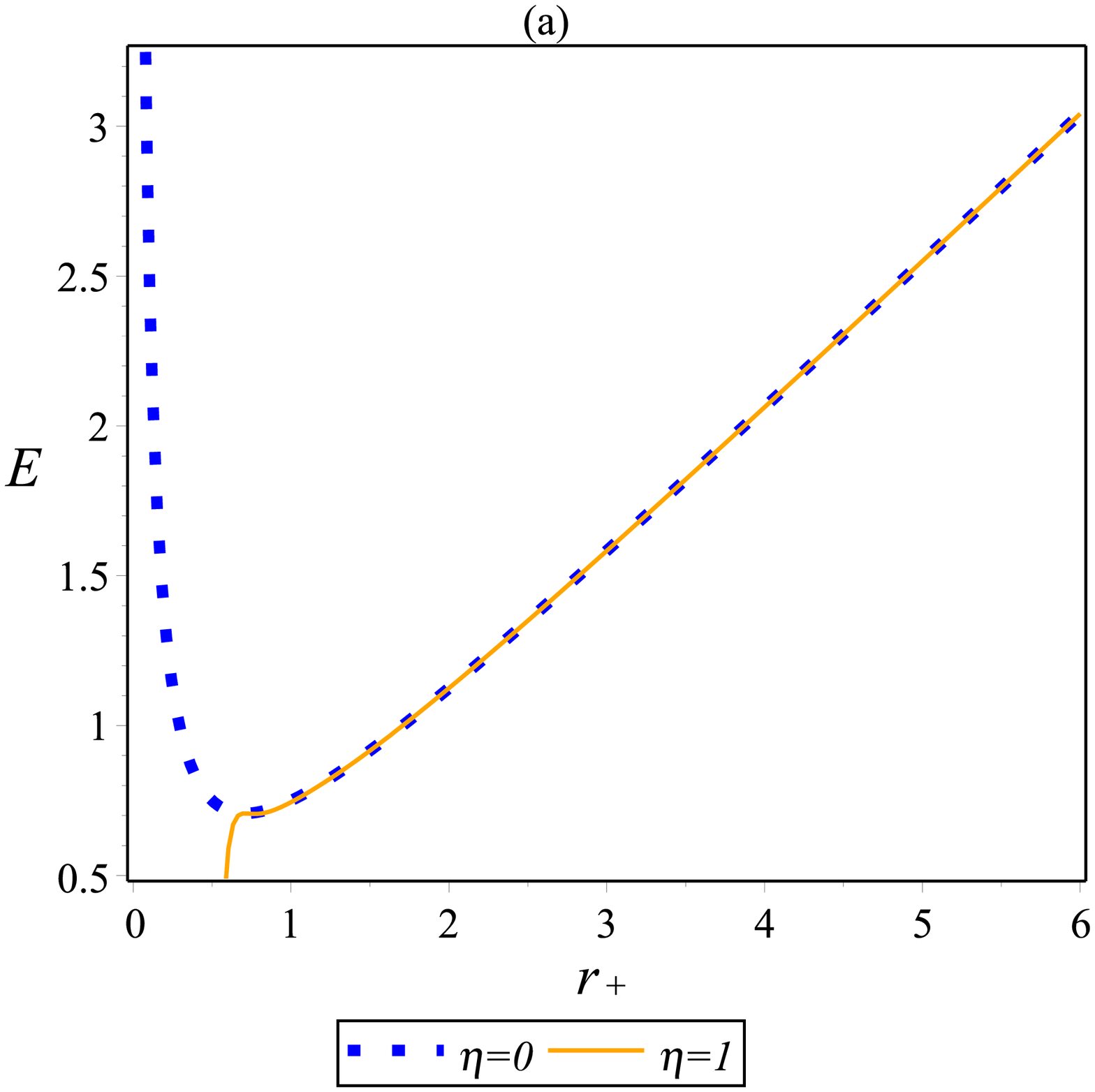}\includegraphics[width=55 mm]{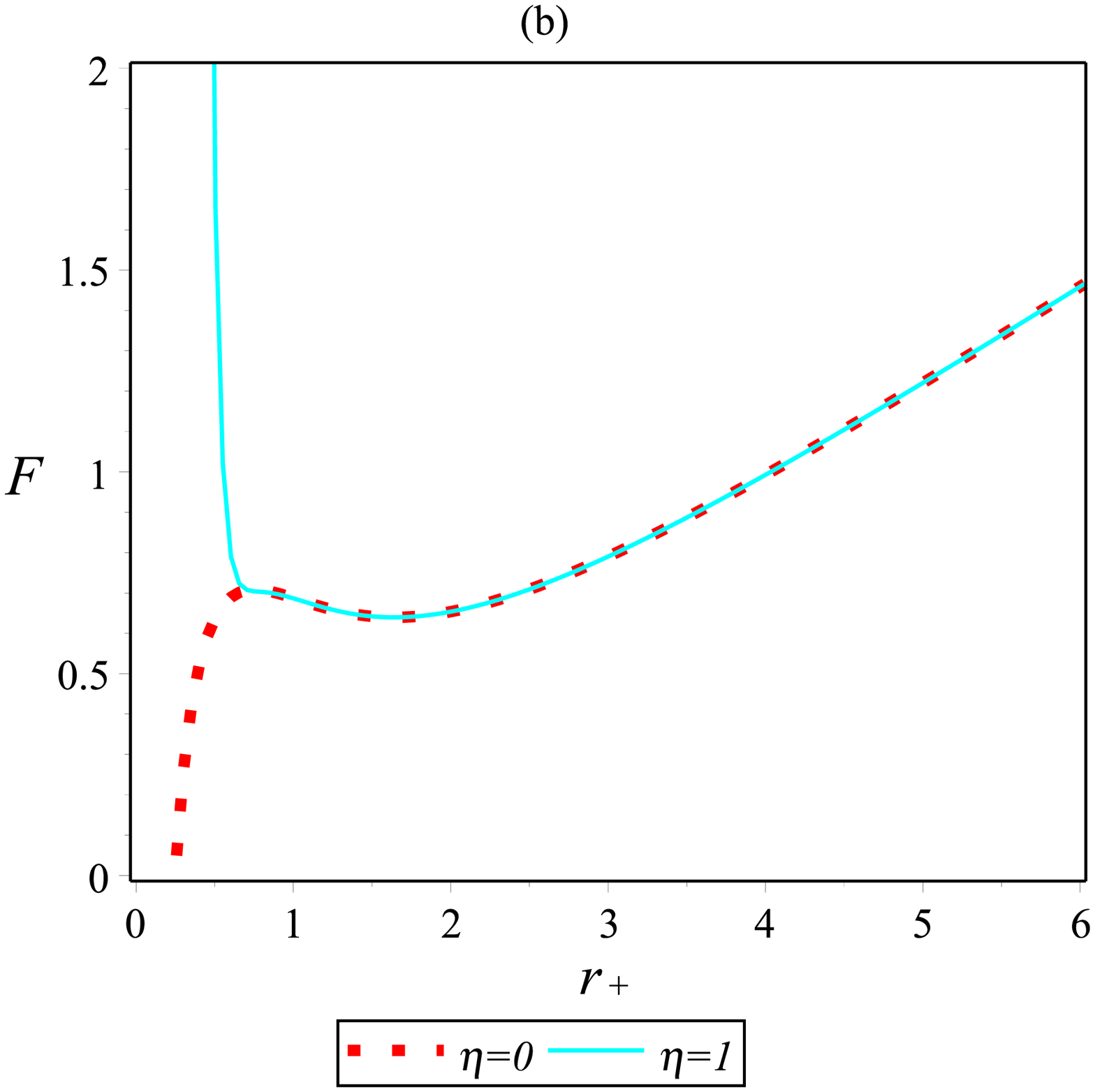}
\end{array}$
\end{center}
\caption{(a) Internal energy, and (b) Helmholtz free energy of KS solution with $k=1$ and $\omega=2$.}
\label{fig8}
\end{figure}

Assigning the black hole mass as being identical to the uncorrected internal energy, the first law of black hole thermodynamics (\ref{first law KS}) emerges in its standard form:
\begin{equation}\label{first law KS2}
dm=TdS_{0}.
\end{equation}
Then, we consider a well-known thermodynamics and statistical relation between the entropy and partition function ($Z$) of a canonical
ensemble which is given by (in geometrized units with $k_{B}=1$),
\begin{equation}\label{SZ}
S=\ln{Z}+\frac{E}{T},
\end{equation}
where $E=m+{\mathcal{O}}(\eta)$ (see Eq. (\ref{mass-KS})). Therefore,
\begin{eqnarray}\label{EZ-KS}
E&=&T^{2}\left(\frac{\partial \ln{Z}}{\partial T}\right)_{V}.\nonumber\\
m&=&\left[T^{2}\left(\frac{\partial \ln{Z}}{\partial T}\right)_{V}\right]_{\eta=0}.
\end{eqnarray}
It means that the black hole mass is unaffected by non-perturbative corrections while the internal energy is altered.
Combination of the expressions (\ref{mass-KS}), (\ref{temp-KS}), and (\ref{S-KS}) with (\ref{SZ}) gives us the logarithm of partition function as follows
\begin{equation}\label{lnZ-KS}
\ln{Z}=\frac{2\pi\left[(2\omega r^{2}-1)\ln{(r)}-\omega^{2}r^{4}+\frac{7}{2}\omega r^{2}+1\right]}{\omega(2\omega r^{2}-1)}+\eta e^{-\pi r^{2}}r^{-\frac{2\pi}{\omega}}.
\end{equation}
We should employ the aforesaid partition function at $r=r_{+}$ in order to achieve thermodynamics.
In order to obtain thermodynamics, we should use the above partition function at $r=r_{+}$. In that case ($r=r_{+}$), by using Eqs. (\ref{EZ-KS}) and (\ref{lnZ-KS}), the corrected internal energy is given by
\begin{equation}\label{E-KS}
E=m+\eta\frac{(\omega r_{+}^{2}+1)(2\omega r_{+}^{2}-1)^{2}e^{-\pi r_{+}^{2}}r_{+}^{-\frac{2\pi+\omega}{\omega}}}{4\omega(2\omega^{2}r_{r}^{4}-5\omega r_{+}^{2}-1)}.
\end{equation}
Moreover, using the well-known expression of the Helmholtz free energy:
\begin{equation}\label{F-KS}
F=-T\ln{Z},
\end{equation}
one can obtain
\begin{equation}
F=\frac{(1-2\omega r_{+}^{2})\ln(r_{+})+\omega^{2}r_{+}^{4}+\frac{7}{2}\omega r_{+}^{2}+1}{4\omega r_{+}(1+\omega r_{+}^{2})}
-\eta\frac{(\omega r_{+}^{2}-\frac{1}{2})e^{-\pi r_{+}^{2}}r_{+}^{-\frac{2\pi}{\omega}}}{4\pi r_{+}(\omega r_{+}^{2}+1)}.\label{F-KS-r}
\end{equation}
Fig. \ref{fig8} (a) exhibits the behavior of the internal energy for the KS solution and also shows the effects of non-perturbative corrections. Ignoring the quantum correction (i.e., $\eta=0$), the internal energy reaches a minimum at the event horizon (critical radius). The non-perturbative corrections yield sudden vanishing internal energy near that critical radius. Fig. \ref{fig8} (b) shows the critical behavior of the Helmholtz free energy. The critical behavior of the Helmholtz free energy means that $\frac{dF}{dr_{+}}=0$, while it is not a minimum or maximum but $\frac{d^{2}F}{dr_{+}^{2}}=0$  \big(see the cyan solid line of Fig. \ref{fig8} (b)\big). Those plots highlight the significance of the non-perturbative corrections at small radii.\\

Since there is no dual variable pressure, we can assume that $P=0$:
\begin{equation}\label{dF-KS}
\left(\frac{dF}{dV}\right)_{T}=0,
\end{equation}
which yields
\begin{equation}\label{H-KS}
H=E,
\end{equation}
where $H$ denotes the enthalpy. Namely, in the absence of the non-perturbative corrections, black hole enthalpy is identical to the black hole mass \cite{main}. Hence, the Gibbs free energy is equal to the Helmholtz free energy.

\section{Corrected thermodynamics of LMP solution}\label{LMP}
\begin{figure}[h!]
\begin{center}$
\begin{array}{cccc}
\includegraphics[width=55 mm]{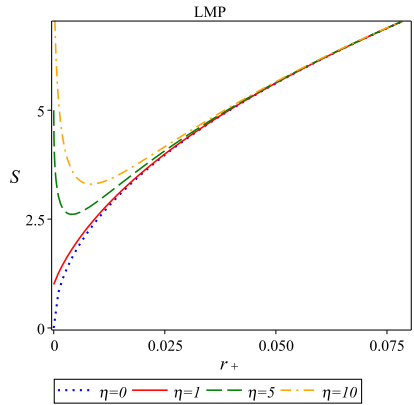}
\end{array}$
\end{center}
\caption{Corrected entropy of LMP solution with $k=1$, $\Lambda_{W}=-1$, and $a=1$.}
\label{fig9}
\end{figure}

\begin{figure}[h!]
\begin{center}$
\begin{array}{cccc}
\includegraphics[width=55 mm]{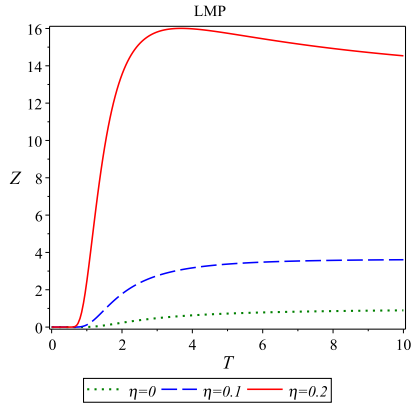}
\end{array}$
\end{center}
\caption{Partition function of LMP solution with $\Lambda_{W}=-1$ and $a=1$.}
\label{fig10}
\end{figure}

According to Ref. \cite{EPJC}, the original entropy of the LMP metric in spherical coordinates ($k=1$) is given by
\begin{equation}\label{s-LMP}
S_{0}=8\pi\sqrt{-\Lambda_{W}r_{+}},
\end{equation}
where $a=1$ is assumed, and $r_{+}$ is the largest root of $f_{LMP}(r)=0$ (see Eq. (\ref{9})). The existence of the cosmological constant implies that there is a pressure which is given by (\ref{pressure}). In that case, the corrected entropy (\ref{S}) can be rewritten as
\begin{equation}\label{S-LMP}
S=8\pi\sqrt{-\Lambda_{W}r_{+}}+\eta e^{-8\pi\sqrt{-\Lambda_{W}r_{+}}}.
\end{equation}
In Fig. \ref{fig9}, the effect of the exponential correction on the black hole entropy is depicted. Thus, the importance of non-perturbative effects at small radii is highlighted. It is clear that the non-perturbative quantum correction increases with the value of the entropy.
As the original black hole entropy is proportional to the event horizon area ($S_{0}=\frac{A}{4}$), then the black hole volume is proportional to $S_{0}r_{+}$. Therefore, one can write the volume as follows
\begin{equation}\label{V}
V=v_{0}S_{0}r_{+}=8\pi v_{0}r_{+}\sqrt{-\Lambda_{W}r_{+}},
\end{equation}
where $v_{0}$ is a constant which can be set to one without loss of generality. The volume can be considered as the conjugate variable of pressure $P$, which is given by (\ref{pressure}).
Also, from Eq. (\ref{temp-LMP}) one can write
\begin{equation}\label{r+LMP}
r_{+}\approx\frac{1}{8\pi T},
\end{equation}
which gives us the following temperature dependent entropy:
\begin{equation}\label{S-T-LMP}
S\approx\sqrt{\frac{8\pi }{T}}(1-\eta)+\frac{4\pi \eta}{T}.
\end{equation}
Now, using Eqs. (\ref{SZ}) and (\ref{EZ-KS}), we can obtain
\begin{equation}\label{izzetLMP}
\ln{Z}=\frac{\sqrt{2}T(\eta-1)\left[\pi \exp\left({\frac{1}{T}}\right)erf\left(\sqrt{\frac{1}{T}}\right)-2\sqrt{\frac{\pi}{T}}\right]+4\pi\eta(1+T)}{T},
\end{equation}
where $erf(x)$ is the error function. In Eq. \eqref{izzetLMP}, the integral constant is simply set to zero without loss of generality. In Fig. \ref{fig10}, the partition function $Z$ is plotted in terms of the temperature. We can see that the non-perturbative corrections increase the value of the partition function at high temperature, while it has no important effect at the low temperature (large radius).

\begin{figure}[h!]
\begin{center}$
\begin{array}{cccc}
\includegraphics[width=55 mm]{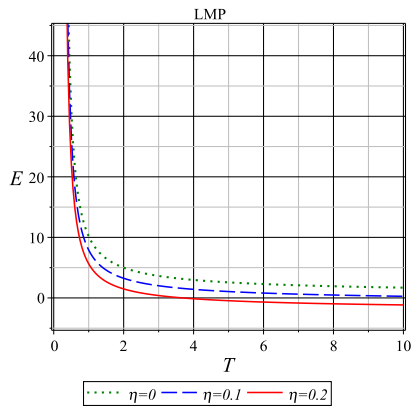}
\end{array}$
\end{center}
\caption{Internal energy of LMP solution with $\Lambda_{W}=-1$ and $a=1$.}
\label{fig11}
\end{figure}

Using the first equation of (\ref{EZ-KS}), one can obtain
\begin{equation}\label{E-LMP}
E=erf\left(\sqrt{\frac{1}{T}}\right)\sqrt{2}\pi\exp\left({\frac{1}{T}}\right)-\eta\pi\left(4+\sqrt{2}erf\left(\sqrt{\frac{1}{T}}\right)\exp\left({\frac{1}{T}}\right)\right).
\end{equation}
In Fig. \ref{fig11}, we can see the typical behavior of the internal energy in terms of the temperature. The figure shows the effects of non-perturbative corrections. As expected, it is clear that at low temperature (large black hole) the non-perturbative quantum correction is negligible.\\

\begin{figure}[h!]
\begin{center}$
\begin{array}{cccc}
\includegraphics[width=55 mm]{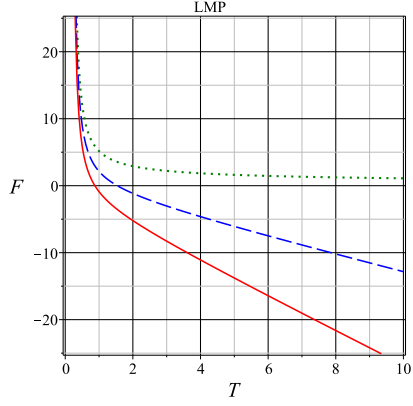}
\end{array}$
\end{center}
\caption{Helmholtz free energy of LMP solution with $\Lambda_{W}=-1$.}
\label{fig12}
\end{figure}

As a next step, using Eq. (\ref{F-KS}), one can obtain the Helmholtz free energy as,
\begin{equation}\label{F-LMP}
F=2\sqrt{2}(\eta-1)\sqrt{\frac{\pi}{T}}-4\pi\eta(1+T)-(\eta-1)erf\left(\sqrt{\frac{1}{T}}\right)\sqrt{2}\pi T\exp\left({\frac{1}{T}}\right).
\end{equation}
In Fig. \ref{fig12}, we can see the behavior of Helmholtz free energy in terms of the temperature. It is completely positive in the absence of non-perturbative correction, which reduces the value of the Helmholtz free energy.

Now, using radius (\ref{r+LMP}) in the expressions of volume (\ref{V}) and pressure (\ref{pressure}), one can obtain the Gibbs free energy via
\begin{equation}\label{G}
G=F+PV,
\end{equation}
which yields,
\begin{equation}\label{G-LMP}
G={\frac {3\,\sqrt {2}}{64} \left( {\frac {1}{\pi \,T}} \right)^{3/2}}+\sqrt {2} T\left(\eta -1\right)\left[
2 \sqrt {{\frac {\pi }{T}}}-\pi \exp\left({\frac{1}{T}}\right)  erf\left(\sqrt{\frac{1}{T}}\right)\right]-4\,\pi \,\eta\, \left( T+1 \right).
\end{equation}
The behavior of the Gibbs free energy is similar to the Helmholtz free energy which means that Gibbs free energy becomes zero in the presence of the non-perturbative correction. It shows a Hawking-Page phase transition, hence we expect some stable and unstable regions in the presence of non-perturbative quantum correction, which should be investigated by specific heat analysis.

\section{Stability and phase transition}\label{stab}
The sign of specific heat tells us some information about the black hole phase. An unstable black hole has negative specific heat, its temperature increases while its mass decreases. The black hole specific heat can be obtained by employing the partition function via the following relation:
\begin{equation}\label{Cv}
C_{v}=T\left[2\left(\frac{d\ln{Z}}{dT}\right)_{V}+T\left(\frac{d^{2}\ln{Z}}{dT^{2}}\right)_{V}\right].
\end{equation}
We are now going to apply it for KS and LMP solutions in the presence of non-perturbative quantum corrections.
\subsection{KS black hole}
Using Eqs. (\ref{Cv}) and (\ref{lnZ-KS}), the specific heat at constant volume for the KS solution is obtained as following
\begin{equation}\label{Cv-KS}
C_{v}=\frac{16\pi\eta}{\omega^{2}}\frac{(\omega r_{+}^{2}+1)^{2}(2\omega r_{+}^{2}-1)}{(2\omega^{2}r_{+}^{4}-5\omega r_{+}^{2}-1)^{3}}
r_{+}^{-\frac{2\pi}{\omega}}\exp\left({-\pi r_{+}^{2}}\right)C_{1}+C_{v0},
\end{equation}
where
\begin{eqnarray}\label{C1}
C_{1}&\equiv&r_{+}^{8}(\pi r_{+}^{2}-\frac{1}{2})\omega^{5}-r_{+}^{6}(\pi r_{+}^{2}-\frac{7}{2})\omega^{4}-\frac{r_{+}^{4}}{4}(17\pi r_{+}^{2}-\frac{15}{2})\omega^{3}\nonumber\\
&-&\frac{5}{2}r_{+}^{2}(\frac{1}{2}\pi r_{+}^{2}-1)\omega^{2}+\frac{1}{4}(5\pi r_{+}^{2}+\frac{1}{2})\omega+\frac{\pi}{4},
\end{eqnarray}
and the uncorrected ($\eta=0$) specific heat is given by
\begin{equation}\label{Cv-KS}
C_{v0}=\frac{2\pi(\omega r_{+}^{2}+1)^{2}(2\omega r_{+}^{2}-1)}{\omega(2\omega^{2}r_{+}^{4}-5\omega r_{+}^{2}-1)}.
\end{equation}

\begin{figure}[h!]
\begin{center}$
\begin{array}{cccc}
\includegraphics[width=55 mm]{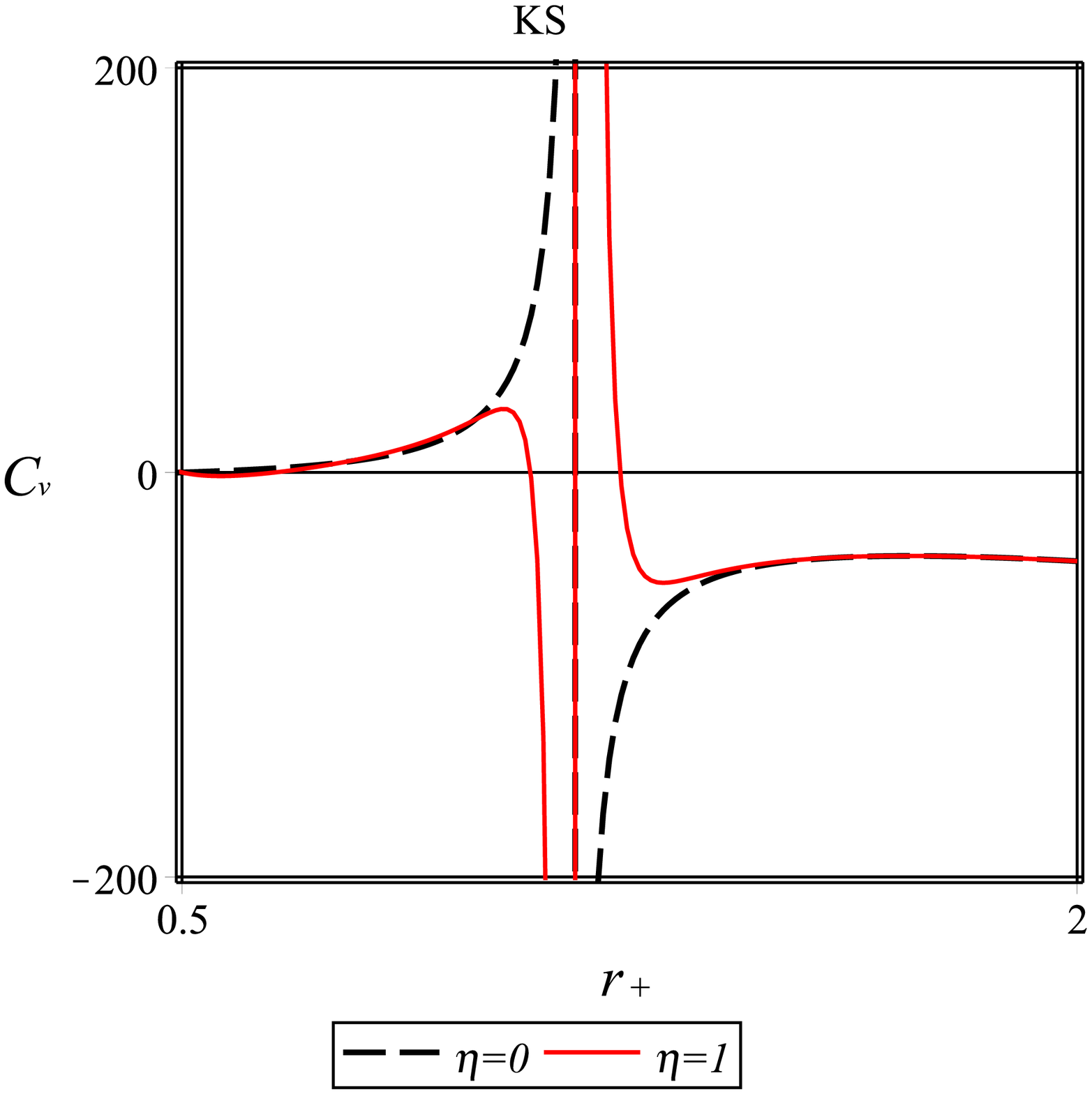}\includegraphics[width=55 mm]{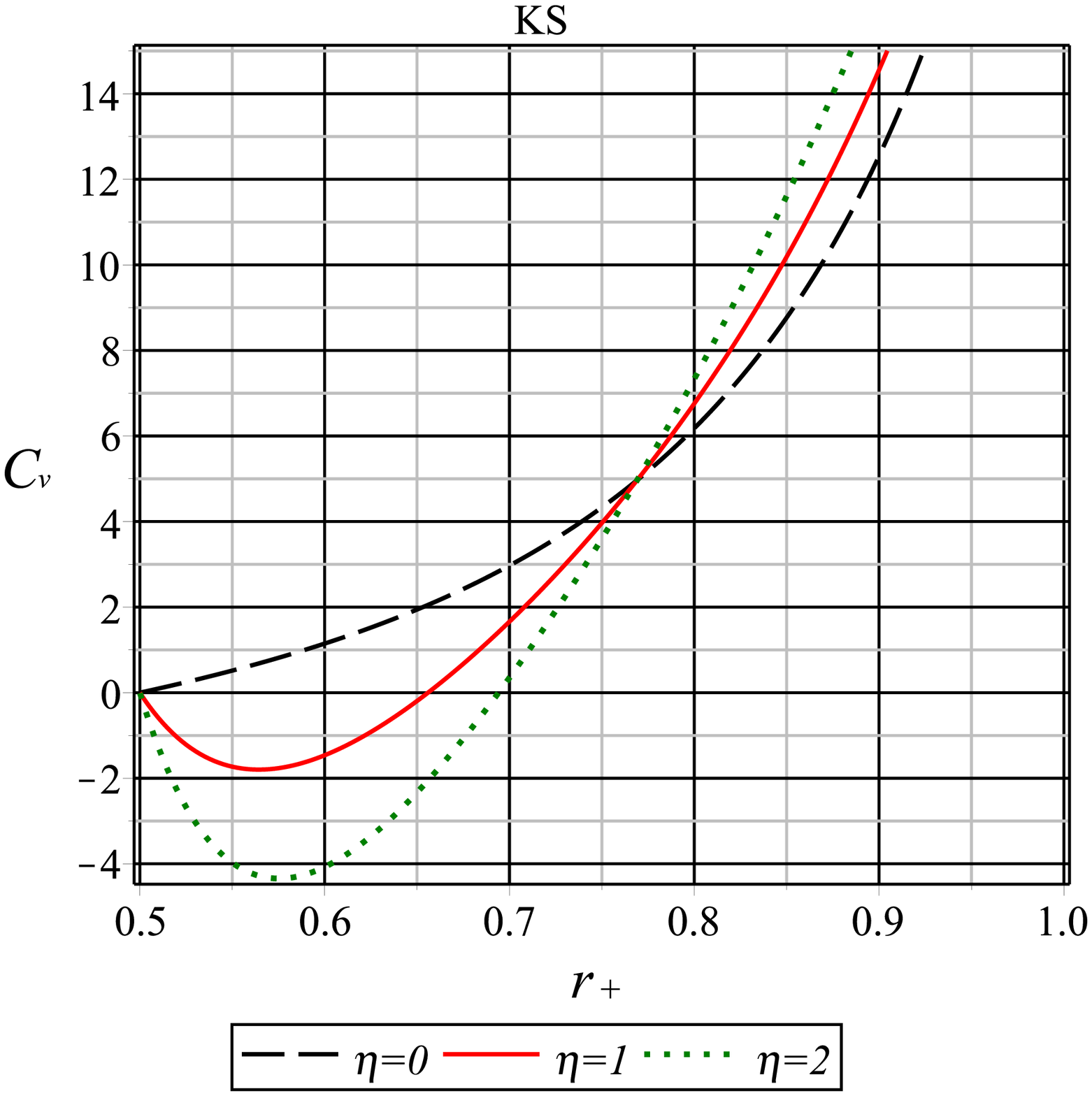}
\end{array}$
\end{center}
\caption{Specific heat in terms of event horizon (radius) of KS solution with $\omega=2$.}
\label{fig13}
\end{figure}

As can be seen from Fig. \ref{fig13}, for the large horizon radius, we find that the non-perturbative correction vanishes. Also, it does not affect the asymptotic point (see left panel of Fig. \ref{fig13}) but it yields some instability at an infinitesimal radius. As we mentioned already, the first order phase transition (asymptotic point) happens at maximum temperature (see Fig. \ref{fig4}). To make the latter remark more obvious, we focus on the smallest radii shown in the right side of Fig. \ref{fig13}. Regarding Eq. (\ref{rm}), we can find that the minimum of horizon radius is $r_{+}=0.5$, so we see that the specific heat reduces to zero at this point. Hence, if we omit the non-perturbative correction, the KS black hole becomes stable at the final stage, when the Hawking radiation ceased. But, in the presence of quantum corrections, the KS black hole turns out to be unstable before stopping of the radiation. This can be interpreted as the second order phase transition which yields the complete black hole evaporation.

\subsection{LMP black hole}
\begin{figure}[h!]
\begin{center}$
\begin{array}{cccc}
\includegraphics[width=55 mm]{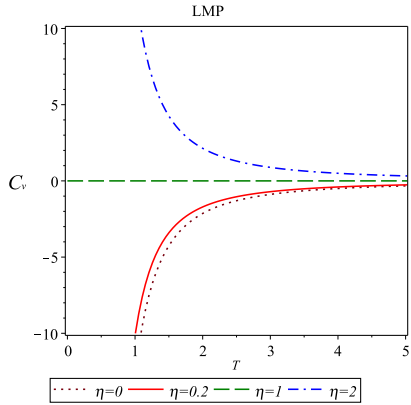}
\end{array}$
\end{center}
\caption{Specific heat in terms of temperature of LMP solution with $\Lambda_{W}=-1$.}
\label{fig14}
\end{figure}

For the LMP black hole, using Eqs. (\ref{izzetLMP}) and (\ref{Cv}), the specific heat at constant volume is obtained as following
\begin{equation}\label{Cv-LMP}
C_{v}=\frac{\sqrt{2}(\eta-1)\left[\sqrt{\pi}+\frac{\pi}{\sqrt{T}}\exp\left({\frac{1}{T}}\right)  erf\left(\sqrt{\frac{1}{T}}\right)\right]}{T^{\frac{3}{2}}}.
\end{equation}
As is clear from Fig. \ref{fig14}, the strength of the non-perturbative correction is important for black hole stability. While the LMP black hole with negligible non-perturbative corrections is completely at unstable phase, however by increasing the quantum effects, it becomes stable. Namely, the specific heat of the LMP solution is negative, while the non-perturbative quantum corrections have a positive contribution to it. There is a special case ($\eta=1$), where these two parts are equal (see the dashed green line of Fig. \ref{fig14}). So the LMP black hole could be completely stable in the presence of non-perturbative quantum corrections. The consequence of such stability is that the LMP black hole can behave as a well-known fluid. In order to find the holographic dual of the LMP black hole, one should study its equation of state. Therefore, we will examine this issue in the next section.

\section{Equation of state}\label{EoS}
\begin{figure}[h!]
\begin{center}$
\begin{array}{cccc}
\includegraphics[width=55 mm]{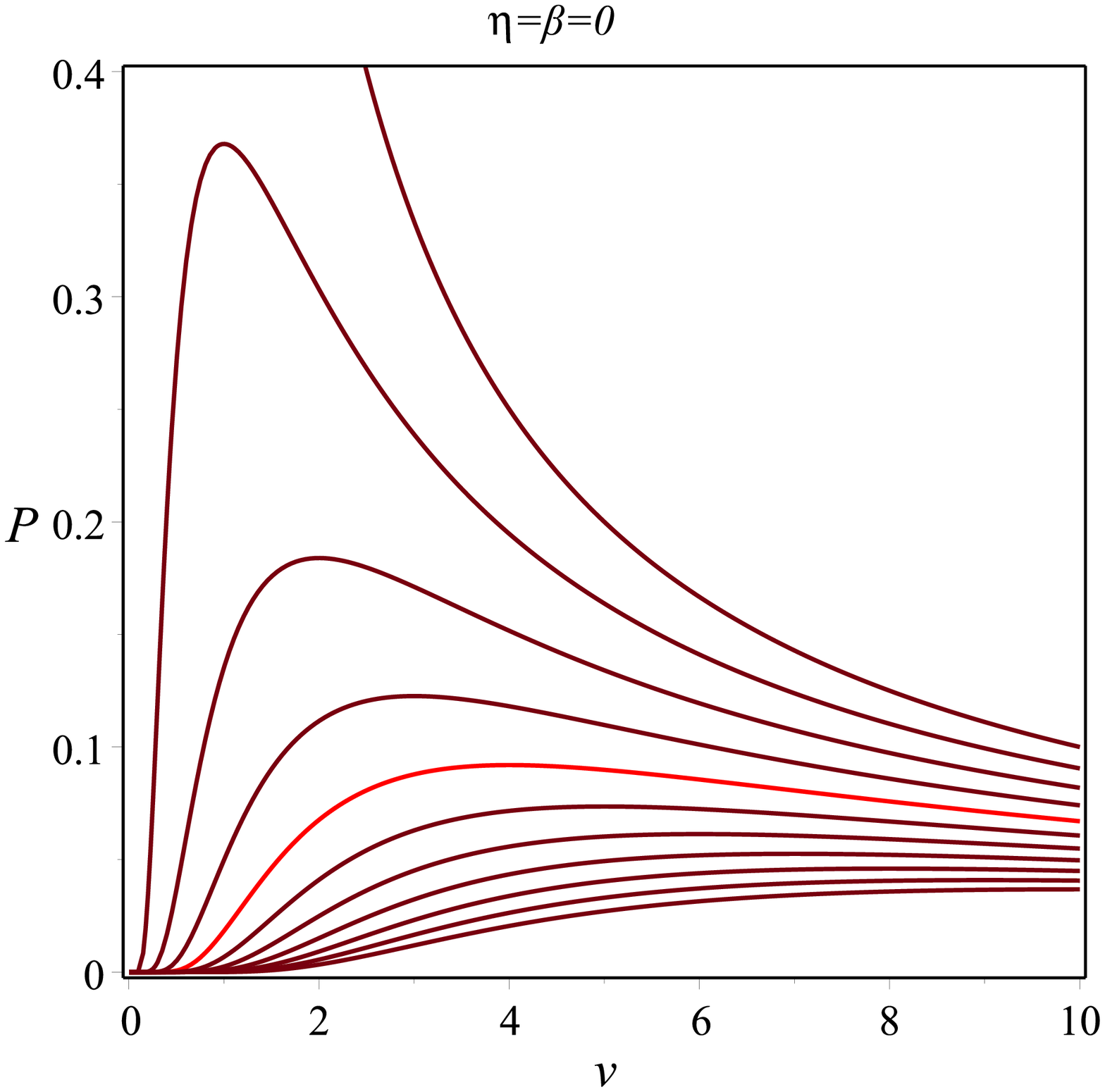}\includegraphics[width=55 mm]{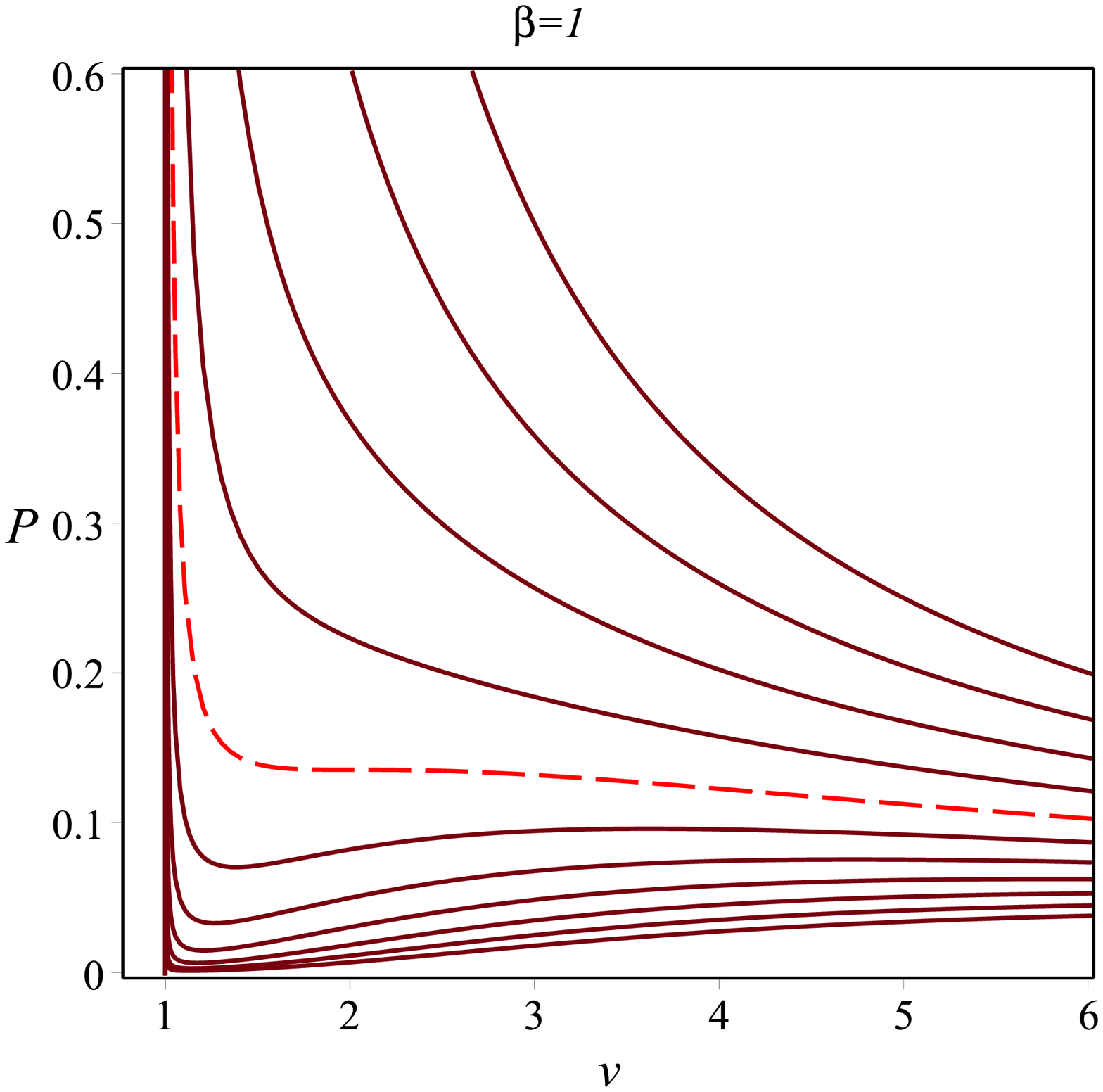}
\end{array}$
\end{center}
\caption{Isotherms for the LMP equation of state with $T=1$. In the left panel $c=0, \cdots, 10$ (uncorrected), while in the right panel $\alpha=0, \cdots, 10$ (corrected) from top to bottom.}
\label{fig15}
\end{figure}

We can study the state equation of LMP solution by using the pressure given by Eq. (\ref{pressure}). According to Eq. (\ref{temp-LMP}), we can obtain $\Lambda_{W}$ as follows
\begin{equation}\label{Lambda}
\Lambda_{W}=\frac{1-8\pi r_{+}T}{5 r_{+}^{2}},
\end{equation}
and thus the pressure given in Eq. (\ref{pressure}) becomes
\begin{equation}\label{P-Lambda}
P=\frac{3}{16\pi}\left[\frac{8\pi T}{5r_{+}}-\frac{1}{5r_{+}^{2}}\right].
\end{equation}
Regarding Eq. (\ref{V}), we define the specific volume \cite{18-0} as following
\begin{equation}\label{specific v}
v=\frac{10}{3}\left((\frac{V}{8\pi})^{2}\frac{1}{-\Lambda_{W}}\right)^{\frac{1}{3}}.
\end{equation}
It means that $10r_{+}=3v$.
Combining Eqs. (\ref{V}), (\ref{specific v}), and (\ref{P-Lambda}), we get
\begin{equation}\label{P-v}
P=\frac{T}{v}-\frac{5}{12\pi v^{2}}.
\end{equation}
It is interesting to compare it with the standard form of the virial expansion, which  expresses the pressure $P$ of a many-particle system in equilibrium as a power series:
\begin{equation}\label{P-v}
\frac{Pv}{T}=1+\frac{B(T)}{v}+\frac{C(T)}{v^{2}}+\cdots,
\end{equation}
where the time-dependent parameters $B(T)$, $C(T)$, \textit{etc.} are called virial coefficients. For the case of LMP solution, it is easy to find that $B(T)=-\frac{5}{12\pi T}$, while the other coefficients are zero. It is analogous with a Van der Waals fluid of point like particles. The Van der Waals equation of state at the first order approximation can be rewritten as
\begin{equation}\label{vdW}
\frac{Pv}{T}=1+\frac{1}{v}\left(b-\frac{c}{T}\right),
\end{equation}
where $b$ denotes the size of particles and $c$ denotes the interaction strength of particles. So, it reduces to the LPM equation of state (\ref{P-v}) if we set $b=0$ and $c=\frac{5}{12\pi}$. Therefore, the LMP black hole behaves like interacting gas of point like particles.\\
The LMP pressure (\ref{P-v}) can be split to the repulsive ($P_{+}$) and attractive ($P_{-}$) pressures as
\begin{equation}\label{PP+-}
P=P_{+}+P_{-},
\end{equation}
where
\begin{eqnarray}\label{P+-}
P_{+}&=&\frac{T}{v},\nonumber\\
P_{-}&=&-\frac{5}{12\pi v^{2}}.
\end{eqnarray}
On the other hand, the right hand side of Eq. (\ref{vdW}) with $b=0$ can be approximated to an exponential function. So, we can write
\begin{equation}\label{B}
Pv=T\exp\left(-\frac{c}{Tv}\right).
\end{equation}
To see the effect of non-perturbative quantum correction, we use Eq. (\ref{S-T-LMP}) which can be reduced to the following relation at $\eta=0$
\begin{equation}\label{S0-T-LMP}
S_{0}\approx\sqrt{\frac{8\pi }{T}}.
\end{equation}
On the other hand, Eq. (\ref{S-T-LMP}) for the infinitesimal $\eta$ can approximate to
\begin{equation}\label{S-T-LMP-app}
S\approx\sqrt{\frac{8\pi }{T}}(1-\eta).
\end{equation}
Now, it is easy to see that Eq. (\ref{S0-T-LMP}) can be modifed to Eq. (\ref{S-T-LMP-app}) by the following transformation:
\begin{equation}\label{Tt}
T\rightarrow\frac{T}{(1-\eta)^2}.
\end{equation}
Thus, we can see the effect of non-perturbative quantum correction by applying the transformation (\ref{Tt}) to Eq. (\ref{B}), which yields
\begin{equation}\label{BTt}
Pv=\frac{T}{(1-\eta)^2}\exp\left(-\frac{c(1-\eta)^2}{Tv}\right).
\end{equation}
Now, if we set $c(1-\eta)^2\equiv \alpha$, we can rewrite Eq. (\ref{BTt}) as
\begin{equation}\label{BB}
Pv(1-\eta)^2=T\exp\left(-\frac{\alpha}{Tv}\right).
\end{equation}
For the infinitesimal $\eta$, we use $(1-\eta)^2\approx1-2\eta$ and set $2\eta v\equiv \beta$. Hence, Eq. (\ref{BB}) is reduced to the following expression
\begin{equation}\label{BBB}
P(v-\beta)=T\exp\left(-\frac{\alpha}{Tv}\right),
\end{equation}
which is indeed the Dieterici equation \cite{Chakraborty:2021enp}. The physical meaning of $\alpha$ is the strength of attractive interactions ($P_{-}$), while the parameter $\beta$ denotes the strength of non-perturbative quantum correction, which is proportional to repulsive pressure ($P_{+}$). In the left panel of Fig. \ref{fig15}, we draw $P-v$ diagram using (\ref{B}), while in the right panel of Fig. \ref{fig15} we have plotted the $P-v$ diagram by using Eq. (\ref{BBB}) to see the effect of non-perturbative quantum correction. The critical point \cite{R} is clear by the red-dashed line of the right plot in Fig. \ref{fig15}. It shows that the Van der Waals equation of state is only possible in the presence of non-perturbative quantum correction. Hence, we can conclude that the quantum LMP black hole is nothing but the holographic dual of a Van der Waals fluid.

\section{Conclusion}\label{con}
In this paper, we considered the KS and LMP solutions of Ho\v{r}ava-Lifshitz black hole to study the effect of the non-perturbative correction on the thermodynamics quantities. In fact,  the KS solution was a black hole with zero cosmological constant. We first studied the horizon structure and the first law of black hole thermodynamics. We then found that the exponential correction of the black hole entropy affects the thermodynamics quantities of the KS solution at small radii. Next, we showed that the specific heat analysis of the KS solution is stable by neglecting the quantum corrections while the non-perturbative corrections make the KS black hole unstable at a small horizon radius.\\
Subsequently, we considered the LMP solution, which is the Ho\v{r}ava-Lifshitz black hole in the presence of the negative cosmological constant. We also discussed the first law of thermodynamics for this black hole solution. We obtained the modified partition function of the LMP solution due to the exponential correction. Then, we used it to compute the thermodynamical potentials like the Helmholtz and Gibbs free energies. Contrary to the KS solution, we found that the LMP solution is completely unstable in the absence of exponential correction, which is due to the negative cosmological constant. However, the existence of the non-perturbative quantum corrections can thermodynamically make it stable. We also derived the virial expansion \cite{Astesiano:2022gph} of LMP black hole pressure and revealed that the obtained result behaves like the Van der Waals equation of state in the presence of non-perturbative quantum corrections. We plotted the isotherms for the LMP equation of state and showed the critical point exists in the presence of exponential correction.\\
Finally, we showed that the non-perturbative quantum effects correct the black hole entropy with an exponential term. This is happened by the modification of the angular-dependent part of the metric. Hence, it has no effect on the metric function $f(r)$. Therefore, the horizon radius and black hole temperature are kept unchanged. However, there are other quantum gravity theories in the literature where the temperature, location of the horizon, and the entropy are modified \cite{1003.5357}. In summary, we concluded the following outcome from this study: the black hole entropy gets correction but the horizon radius and temperature do not change. Then, we used the general statistical relation between the partition function and the entropy to obtain the quantum corrected or the so-called partition function. Thus, we used the modified partition function to study the modified thermodynamics of Ho\v{r}ava-Lifshitz black hole. The method used in this paper can be extended to other black hole solutions like STU black hole \cite{STU}, G\"{o}del black holes \cite{Godel}, and Hayward black hole \cite{Hay}, which are among our near-future work plans. Such studies can increase our knowledge about quantum black holes and help us to understand why these black holes are never seen experimentally for example in LHC.

\section*{Acknowledgments}

While B. Pourhassan thanks to Iran Science Elites Federation (Tehran), \.{I}. Sakall{\i} gratefully acknowledge the contributions of TUBITAK and SCOAP3.


\begin{thebibliography}{99}
\bibitem{0001}
J. M. Bardeen, B. Carter and S. W. Hawking, "The four laws of black hole mechanics", Commun. Math. Phys. 31 (1973) 161-170. DOI: 10.1007/BF01645742
\bibitem{0002}
J. D. Bekenstein, "Black holes and the second law", Lett. Nuovo Cim. 4 (1972) 737-740. DOI: 10.1007/BF02757029
\bibitem{0003}
J. D. Bekenstein, "Black Holes and Entropy", Phys. Rev. D 7 (1973) 2333. DOI: 10.1103/PhysRevD.7.2333
\bibitem{0004}
S. Sarkar, "Black Hole Thermodynamics: General Relativity and Beyond", Gen. Rel. and Grav. 51 (2019) 63. DOI: 10.1007/s10714-019-2545-y
\bibitem{R16}
H. Ghaffarnejad, E. Ghasemi, "Magnetic charge effects on thermodynamic phase transition of modified Anti de Sitter Ay\'{o}n-Beato-Garc\'{i}a Black Holes with five parameters", JHAP 3 (2022) 47-56. DOI: 10.22128/jhap.2022.524.1022
\bibitem{0005}
M. Moussa, "Schwarzschild Black Hole Thermodynamics and Generalized Uncertainty Principle", Int. J. Theor. Phys. 60 (2021) 994-1007.
\bibitem{0006}
D. A. Gomes, R. V. Maluf, C. A. S. Almeida, "Thermodynamics of Schwarzschild-like black holes in modified gravity models", Annals Phys. 418 (2020) 168198. DOI: 10.1016/j.aop.2020.168198
\bibitem{0007}
K. Ghaderi, B. Malakolkalami, "Thermodynamics of the Schwarzschild and the Reissner-Nordstr\"{o}m black holes with quintessence", Nucl. Phys. B 903 (2016) 10-18. DOI: 10.1016/j.nuclphysb.2015.11.019
\bibitem{0008}
V. G. Czinner, H. Iguchi, "Thermodynamics, stability and Hawking-Page transition of Kerr black holes from R\'{e}nyi statistics", Eur. Phys. J. C 77 (2017) 892. DOI: 10.1140/epjc/s10052-017-5453-x
\bibitem{0009}
O. Ruiz, U. Molina, and P. Viloria, "Thermodynamic analysis of Kerr-Newman black holes", Journal of Physics: Conf. Series 1219 (2019) 012016. DOI: 10.1088/1742-6596/1219/1/012016
\bibitem{AdS}
Z. Gao, L. Zhao, "Restricted phase space thermodynamics for AdS black holes via holography", Class. Quant. Grav. 39 (2022) 075019. DOI: 10.1088/1361-6382/ac566c
\bibitem{00010}
V. E. Hubeny, D. Marolf, M. Rangamani, "Hawking radiation from AdS black holes", Class. Quantum Grav. 27 (2010) 095018. DOI: 10.1088/0264-9381/27/9/095018
\bibitem{00011}
B. Pourhassan, J. Sadeghi and A. Chatrabhuti, "$AdS_{5}$ black hole at N=2 supergravity"  Indian J. Phys. 87 (2013) 691-698. DOI: 10.1007/s12648-013-0287-3
\bibitem{00012}
P. Wang, H. Wu, H. Yang, "Thermodynamic geometry of AdS black holes and black holes in a cavity", Eur. Phys. J. C 80 (2020) 216. DOI: 10.1140/epjc/s10052-020-7776-2
\bibitem{00013}
H. Lu, Y. Pang, C.N. Pope, "AdS Dyonic Black Hole and its Thermodynamics", JHEP 1311 (033) 2013. DOI: 10.1007/JHEP11(2013)033
\bibitem{00014}
C-H. Wu, D-C. Zou, and Y. Wang, "P-V Criticality of Born-Infeld AdS Black Holes Surrounded by Quintessence", Commun. Theor. Phys. 70 (2018) 459. DOI: 10.1088/0253-6102/70/4/459
\bibitem{00015}
A. Adams, C. M. Brown, O. DeWolfe, C. Rosen, "Charged Schrodinger Black Holes", Phys.Rev.D 80 (2009) 125018. DOI: 10.1103/PhysRevD.80.125018
\bibitem{00016}
J. Sadeghi , B. Pourhassan  and  F. Pourasadollah, "Thermodynamics of Schrodinger black holes with hyperscaling violation" Physics Letters B 720 (2013) 244. DOI: 10.1016/j.physletb.2013.02.011
\bibitem{00017}
D. Sudarsky, "A Schrodinger Black Hole and Its Entropy", Mod. Phys. Lett. A  17 (2002) 1047-1057. DOI: 10.1142/S0217732302006928
\bibitem{00018}
J. E. Wang, E. Greenwood, D. Stojkovic, "Schrodinger formalism, black hole horizons and singularity behavior", Phys. Rev. D 80 (2009) 124027. DOI: 10.1103/PhysRevD.80.124027
\bibitem{00019}
J. Sadeghi, B. Pourhassan and A. Asadi, "Thermodynamics of string black hole with hyperscaling violation" Eur. Phys. J. C 74 (2014) 2680. DOI: 10.1140/epjc/s10052-013-2680-7
\bibitem{00020}
C. Eling, Y. Oz, "Ho\v{r}ava-Lifshitz black hole hydrodynamics", JHEP 2014 (2014) 67. DOI: 10.1007/JHEP11(2014)067
\bibitem{00021}
D. Bazeia, F.A. Brito, F.G. Costa, "Two Dimensional Ho\v{r}ava-Lifshitz Black Hole Solutions",  Phys. Rev. D 91 (2015) 044026. DOI: 10.1103/PhysRevD.91.044026
\bibitem{00022}
E. Kiritsis, G. Kofinas, "On Ho\v{r}ava-Lifshitz Black Holes", JHEP 2010 (2010) 122. DOI: 10.1007/JHEP01(2010)122
\bibitem{00023}
P. Ho\v{r}ava, "Membranes at Quantum Criticality", JHEP 0903 (2009) 020. DOI: 10.1088/1126-6708/2009/03/020
\bibitem{00024}
P. Ho\v{r}ava, "Spectral Dimension of the Universe in Quantum Gravity at a Lifshitz Point", Phys. Rev. Lett. 102 (2009) 161301. DOI: 10.1103/PhysRevLett.102.161301
\bibitem{00025}
G. Calcagni, "Cosmology of the Lifshitz universe" JHEP 0909 (2009) 112. DOI: 10.1088/1126-6708/2009/09/112
\bibitem{00026}
R.G. Cai, L.M. Cao, N. Ohta, "Topological black holes in Ho\v{r}ava-Lifshitz gravity", Phys. Rev. D 80 (2009) 024003. DOI: 10.1103/PhysRevD.80.024003
\bibitem{00027}
R.G. Cai, Y. Liu, Y.W. Sun, "On the $z = 4$ Ho\v{r}ava-Lifshitz gravity", JHEP 0906 (2009) 010. DOI: 10.1088/1126-6708/2009/06/010
\bibitem{00028}
J. Sadeghi  and B. Pourhassan, K. Jafarzadeh, E. Reisi and M. Rostami, "Massless Fermion Quasinormal Modes in the Ho\v{r}ava-Lifshitz Background" Can. J. Phys. 91 (2013) 251-255. DOI: 10.1139/cjp-2012-0552
\bibitem{000281}
H. Xu, Y. C. Ong, "Black hole evaporation in Ho\v{r}ava-Lifshitz gravity", Eur. Phys. J. C 80 (2020) 679. DOI: 10.1140/epjc/s10052-020-8249-3
\bibitem{00029}
R.B. Mann, S.N. Solodukhin, "Universality of quantum entropy for extreme black holes", Nucl. Phys. B 523 (1998)
293-307. DOI: 10.1016/S0550-3213(98)00094-7
\bibitem{00030}
J. Sadeghi, B. Pourhassan, and F. Rahimi, "Logarithmic corrections to charged hairy black hole in (2+1) dimensions", Can. J. Phys. 92 (2014) 1638. DOI: 10.1139/cjp-2014-0229
\bibitem{00031}
A. J. M. Medved, G. Kunstatter, "One-loop corrected thermodynamics of the extremal and nonextremal spinning Banados-Teitelboim-Zanelli black hole", Phys. Rev. D 63 (2001) 104005. DOI: 10.1103/PhysRevD.63.104005
\bibitem{00032}
J. Sadeghi, B. Pourhassan, M. Rostami, "P-V criticality of logarithm-corrected dyonic charged AdS black holes", Phys. Rev. D 94 (2016) 064006. DOI: 10.1103/PhysRevD.94.064006
\bibitem{JHAP}
S. Upadhyay, N. Islam, P. Ganai, "A modified thermodynamics of rotating and charged BTZ black hole", Journal of Holography Applications in Physics 2 (2022) 25-48. DOI: 10.22128/jhap.2021.454.1004
\bibitem{R1}
A. Ghosh, S. Mukherji, C. Bhamidipati, "Novel logarithmic corrections to black hole entropy" [arxiv: 2104.05388]
\bibitem{R2}
Y. H. Khan, S. Upadhyay, P. A. Ganai, "Stability of remnants of Bardeen regular black holes in presence of thermal fluctuations", Mod. Phys. Lett. A 36 (2021) 2130023. DOI: 10.1142/S0217732321300238
\bibitem{R3}
B. Pourhassan, S. Upadhyay, "Perturbed thermodynamics of charged black hole solution in Rastall theory", Eur. Phys. J. Plus 136 (2021) 311. DOI: 10.1140/epjp/s13360-021-01271-9
\bibitem{R4}
M. Dehghani, "Thermal fluctuations of AdS black holes in three-dimensional rainbow gravity", Physics Letters B 793 (2019) 234-239. DOI: 10.1016/j.physletb.2019.04.058
\bibitem{R5}
Y. H. Khan, P. A. Ganai, S. Upadhyay, "Quantum-corrected thermodynamics and P-V criticality of self-gravitating Skyrmion black holes", Prog. Theor. Exp. Phys. 2020 (2020) 103B06. DOI: 10.1093/ptep/ptaa135
\bibitem{R6}
N.-ul Islam, P. A. Ganai, S. Upadhyay, "Thermal fluctuations to thermodynamics of non-rotating BTZ black hole", Prog. Theor. Exp. Phys. 2012 (1019) 103B06. DOI: 10.48550/arXiv.1811.05313
\bibitem{R7}
Q. Ama-Tul-Mughania, A. Waseemb, W. us Salamc, A. Jawad, "Greybody factor and thermal fluctuations of rotating regular black hole bounded by PFDM", Chinese Journal of Physics 77 (2022) 2213-2227. DOI: 10.1016/j.cjph.2021.11.024
\bibitem{R8}
S. Upadhyay, B. Pourhassan, "Logarithmic corrected Van der Waals black holes in higher dimensional AdS space", Prog. Theor. Exp. Phys. 2019 (2019) 013B03. DOI: 10.1093/ptep/pty145
\bibitem{R9}
S. Upadhyay, "Leading-order corrections to charged rotating AdS black holes thermodynamics", Gen. Rel. Grav. 50 (2018) 128. DOI: 10.1007/s10714-018-2459-0
\bibitem{R10}
S. Upadhyay, "Quantum corrections to thermodynamics of quasitopological black holes", Physics Letters B 775 (2017) 130-139. DOI: 10.1016/j.physletb.2017.10.059
\bibitem{R11}
B. Pourhassan, M. Faizal, S. Upadhyay, L. Al Asfar "Thermal Fluctuations in a Hyperscaling Violation Background", Eur. Phys. J. C 77 (2017) 555. DOI: 10.1140/epjc/s10052-017-5125-x
\bibitem{R12}
A. Jawad, M. U. Shahzad, "Effects of Thermal Fluctuations on Non-minimal Regular Magnetic Black Hole", Eur. Phys. J. C 77 (2017) 349. DOI: 10.1140/epjc/s10052-017-4914-6
\bibitem{R13}
S. Saghafi; K. Nozari, "Shadow behavior of the quantum-corrected Schwarzschild black hole immersed in holographic quintessence", Journal of Holography Applications in Physics 3 (2022) 31-38. DOI: 10.22128/jhap.2022.515.1019
\bibitem{R14}
G. Gour, A.J.M. Medved, "Thermal Fluctuations and Black Hole Entropy", Class. Quant. Grav. 20 (2003) 3307-3326. DOI: 10.1088/0264-9381/20/15/303
\bibitem{R15}
A. Desai, S. Sharma, P. A. Ganai, "Quantum fluctuations of a regular charged black hole in massive gravity" [arxiv: 2207.00593]
\bibitem{00033}
B. Pourhassan, K. Kokabi, S. Rangyan, "Thermodynamics of higher dimensional black holes with higher order thermal fluctuations", Gen. Relativ. Gravit 49 (2017) 144. DOI: 10.1007/s10714-017-2315-7
\bibitem{00034}
B. Pourhassan, H. Farahani, S. Upadhyay, "Thermodynamics of Higher Order Entropy Corrected Schwarzschild-Beltrami-de Sitter Black Hole", Int. J. Mod. Phys. A 34 (2019) 1950158. DOI: 10.1142/S0217751X19501586
\bibitem{higher001}
S. S. More. "Higher Order Corrections to Black Hole Entropy", Class. Quantum Grav. 22 (2005) 4129-4140. DOI: 10.1088/0264-9381/22/19/021
\bibitem{00035}
B. Pourhassan, M. Dehghani, M. Faizal, S. Dey, "Non-Pertubative Quantum Corrections to a Born-Infeld Black Hole and its Information Geometry", Class. Quantum Grav. 38 (2021) 105001. DOI: 10.1088/1361-6382/abdf6f
\bibitem{2010.03946}
B. Pourhassan, "Exponential corrected thermodynamics of black holes", J. Stat. Mech. (2021) 073102. DOI: 10.1088/1742-5468/ac0f6a
\bibitem{expJHEP00}
B. Pourhassan, S. S. Wani, S. Soroushfar, M. Faizal, "Quantum Work and Information Geometry of a Quantum Myers-Perry Black Hole, J. High Energ. Phys. 2021 (2021) 27. DOI: 10.1007/JHEP10(2021)027
\bibitem{log5}
A. Dabholkar, J. Gomes and S. Murthy, "Nonperturbative black hole entropy and Kloosterman sums", JHEP 1503 (2015) 074. DOI: 10.1007/JHEP03(2015)074
\bibitem{test003}
B. Pourhassan, M. Faizal, S. Capozziello, "Testing Quantum Gravity through Dumb Holes", Annals of Physics 377 (2017) 108. DOI: 10.1016/j.aop.2016.11.014
\bibitem{expJHEP01}
B. Pourhassan, H. Aounallah,, M. Faizal, S. Upadhyay, S. Soroushfar, Y. O. Aitenov, S. S. Wani, "Quantum Thermodynamics of an M2-M5 Brane System", JHEP 05 (2022) 030. DOI: 10.1007/JHEP05(2022)030
\bibitem{expJHEP02}
B. Pourhassan, M. Faizal, "Quantum Corrections to the Thermodynamics of Black Branes", J. High Energ. Phys. 2021 (2021) 50. DOI: 10.1007/JHEP10(2021)050
\bibitem{log0}
R. K. Kaul, P. Majumdar, "Logarithmic correction to the Bekenstein-Hawking entropy", Phys. Rev. Lett. 84 (2000) 5255. DOI: 10.1103/PhysRevLett.84.5255
\bibitem{22}
S. Upadhyay, "Thermodynamics and galactic clustering with a modified gravitational potential", Phys. Rev. D 95 (2017) 043008. DOI: 10.1103/PhysRevD.95.043008
\bibitem{P1}
A. Iorio, G. Lambiase, P. Pais, F. Scardigli, "Generalized Uncertainty Principle in three-dimensional
gravity and the BTZ black hole", Phys. Rev. D 101 (2020) 105002. DOI: 10.1103/PhysRevD.101.105002
\bibitem{higher3}
B. Pourhassan, K. Kokabi, Z. Sabery, "Higher order corrected thermodynamics and statistics of Kerr-Newman-Godel black hole",  Annals of Physics 399 (2018) 181. DOI: 10.1016/j.aop.2018.10.011
\bibitem{2007.15401}
A. Chatterjee, A. Ghosh, "Exponential corrections to black hole entropy", Phys. Rev. Lett. 125 (2020) 041302. DOI: 10.1103/PhysRevLett.125.041302
\bibitem{EPL}
B. Pourhassan, M. Faizal, "Thermal Fluctuations in a Charged AdS Black Hole", EPL 111 (2015) 40006. DOI: 10.1209/0295-5075/111/40006
\bibitem{NPB}
B. Pourhassan, M. Faizal, "Thermodynamics of a Sufficient Small Singly Spinning Kerr-AdS Black Hole", Nuclear Physics B 913 (2016) 834-851. DOI: 10.1016/j.nuclphysb.2016.10.013
\bibitem{Sen001}
N. Banerjee, D. P. Jatkar, A. Sen. "Asymptotic Expansion of the N=4 Dyon Degeneracy", JHEP 0905 (2009) 121. DOI: 10.1088/1126-6708/2009/05/121
\bibitem{e1}
J. Sadeghi,  M.R. Setare,  B. Pourhassan, "Two Dimensional Black  Hole Entropy", Eur. Phys. J. C 53 (2008) 95-97. DOI: 10.1140/epjc/s10052-007-0435-z

\bibitem{12}
H. Lu, J. Mei, C.N. Pope, "Solutions to Ho\v{r}ava Gravity", Phys. Rev. Lett. 103 (2009) 091301. DOI: 10.1103/PhysRevLett.103.091301
\bibitem{EPJC}
B. Pourhassan, "PV criticality of the second order quantum corrected Ho\v{r}ava-Lifshitz black hole", Eur. Phys. J. C 79 (2019) 740. DOI: 10.1140/epjc/s10052-019-7257-7
\bibitem{13}
M-I. Park, "The Black Hole and Cosmological Solutions in IR modified Ho\v{r}ava Gravity", J. High Energy Phys. 0909 (2009) 123. DOI: 10.1088/1126-6708/2009/09/123
\bibitem{AS}
R. Biswas, and S. Chakraborty, "Black Hole Thermodynamics in Ho\v{r}ava-Lifshitz Gravity and the Related Geometry",
Astrophys. Space Sci. 332 (2011) 193-199. DOI: 10.1007/s10509-010-0504-x
\bibitem{14}
H. W. Lee, Y-W. Kim, Y. S. Myung, "Slowly rotating black holes in the Ho\v{r}ava-Lifshitz gravity", Eur. Phys. J. C 70 (2010) 367-371. DOI: 10.1140/epjc/s10052-010-1463-7
\bibitem{15}
E. Kiritsis, G. Kofinas, "Ho\v{r}ava-Lifshitz Cosmology", Nucl. Phys. B821 (2009) 467-480. DOI: 10.1016/j.nuclphysb.2009.05.005
\bibitem{16}
J. Sadeghi, B. Pourhassan, "Particle acceleration in Ho\v{r}ava-Lifshitz black holes", Eur. Phys. J. C 72 (2011) 1984. DOI: 10.1140/epjc/s10052-012-1984-3
\bibitem{17-1}
D. Y. Chen, H. Yang, X. T. Zu, "Hawking radiation of black holes in the $z=4$
Horava-Lifshitz gravity", Phys. Lett. B681 (2009) 463-468. DOI: 10.1016/j.physletb.2009.10.065
\bibitem{wald}
R.M. Wald, "General Relativity" (The University of Chicago Press, Chicago, 1984)
\bibitem{19}
A. Kehagias and K. Sfetsos, "The black hole and FRW geometries of non-relativistic gravity", Physics Letters B 678 (2009) 123-126. DOI: 10.1016/j.physletb.2009.06.019
\bibitem{18-0}
N. Altamirano, D. Kubiznak, R. B. Mann, Z. Sherkatghanad, "Thermodynamics of rotating black holes and black rings: phase transitions and
thermodynamic volume", Galaxies 2 (2014) 89-159. DOI: 10.3390/galaxies2010089
\bibitem{main}
J. Sadeghi, K. Jafarzade, and B. Pourhassan, "Thermodynamical Quantities of Ho\v{r}ava-Lifshitz Black Hole", Int. J. Theor. Phys. 51 (2012) 3891-3902. DOI: 10.1007/s10773-012-1281-9
\bibitem{NPB2}
B. Pourhassan, S. Upadhyay, H. Saadat, H. Farahani, "Quantum gravity effects on Ho\v{r}ava-Lifshitz black hole", Nucl. Phys. B 928 (2018) 415-434. DOI: 10.1016/j.nuclphysb.2018.01.018
\bibitem{Chakraborty:2021enp}
S.~Chakraborty and D.~Gregoris, "Cosmological evolution with quadratic gravity and nonideal fluids",
Eur. Phys. J. C 81 (2021) 944. DOI: 10.1140/epjc/s10052-021-09697-2
\bibitem{R}
S. Upadhyay, B. Pourhassan, H. Farahani, "P-V criticality of first-order entropy corrected AdS black holes in massive gravity", Phys. Rev. D 95 (2017) 106014. DOI: 10.1103/PhysRevD.95.106014
\bibitem{Astesiano:2022gph}
D.~Astesiano, "Rigid rotation in GR and a generalization of the virial theorem for gravitomagnetism", Gen. Rel. Grav. 54 (2022) 63. DOI: 10.1007/s10714-022-02947-y
\bibitem{1003.5357}
R. C. Myers, B. Robinson, "Black Holes in Quasi-topological Gravity", JHEP 1008 (2010) 067. DOI: 10.1007/JHEP08(2010)067
\bibitem{STU}
B. Pourhassan, M. Faizal, "The lower bound violation of shear viscosity to entropy ratio due to logarithmic correction in STU model", Eur. Phys. J. C 77 (2017) 96. DOI: 10.1140/epjc/s10052-017-4665-4
\bibitem{Godel}
A. Pourdarvish, J. Sadeghi, H. Farahani, and B.Pourhassan, "Thermodynamics and Statistics of Godel Black Hole with Logarithmic Correction" Int. J. Theor. Phys. 52 (2013) 3560-3563. DOI: 10.1007/s10773-013-1658-4
\bibitem{Hay}
B. Pourhassan, M. Faizal, and U. Debnath "Effects of Thermal Fluctuations on the Thermodynamics of Modified Hayward Black Hole", Eur. Phys. J. C 76 (2016) 145. DOI: 10.1140/epjc/s10052-016-3998-8
\end{thebibliography}
\end{document}